\documentclass[%
 aip,
 amsmath,amssymb,
 reprint,%
]{revtex4-1}

\usepackage{graphicx}
\usepackage{dcolumn}
\usepackage{bm}
\usepackage{adjustbox}
\usepackage[utf8]{inputenc}
\usepackage[T1]{fontenc}
\usepackage{mathptmx}
\usepackage{xcolor}
\usepackage{hyperref}
\hypersetup{
	colorlinks   = true,
	citecolor    = blue,
	linkcolor = blue,
    urlcolor = blue
}

\begin{document}

\preprint{AIP/123-QED}

\title{A pathway towards high throughput Quantum Monte Carlo simulations for alloys: A case study of two-dimensional (2D) GaS$_{x}$Se$_{1-x}$}

\author{Daniel Wines}
\affiliation{%
Department of Physics, University of Maryland Baltimore County, Baltimore MD 21250
}%

\author{Kayahan Saritas}%

\affiliation{ 
Department of Applied Physics, Yale University, New Haven CT 06520
}%

\author{Can Ataca}
 \email{ataca@umbc.edu}
\affiliation{%
Department of Physics, University of Maryland Baltimore County, Baltimore MD 21250
}%

\date{\today}

\begin{abstract}

The study of alloys using computational methods has been a difficult task due to the usually unknown stoichiometry and local atomic ordering of the different structures experimentally. In order to combat this, first-principles methods have been coupled with statistical methods such as the Cluster Expansion formalism in order to construct the energy hull diagram, which helps to determine if an alloyed structure can exist in nature. Traditionally, density functional theory (DFT) has been used in such workflows. In this work we propose to use chemically accurate many-body variational Monte Carlo (VMC) and diffusion Monte Carlo (DMC) methods to construct the energy hull diagram of an alloy system, due to the fact that such methods have a weaker dependence on the starting wavefunction and density functional, scale similarly to DFT with the number of electrons, and have had demonstrated success for a variety of materials. To carry out these simulations in a high-throughput manner, we propose a method called \textit{Jastrow sharing}, which involves recycling the optimized Jastrow parameters between alloys with different stoichiometries. We show that this eliminates the need for extra VMC Jastrow optimization calculations and results in a significant computational cost savings (on average 1/4 savings of total computational time). Since it is a novel post-transition metal chalcogenide alloy series that has been synthesized in its few-layer form, we used monolayer GaS$_{x}$Se$_{1-x}$ as a case study for our workflow. By extensively testing our \textit{Jastrow sharing} procedure for monolayer GaS$_{x}$Se$_{1-x}$ and quantifying the cost savings, we demonstrate how a pathway towards chemically accurate high-throughput simulations of alloys can be achieved using many-body VMC and DMC methods.

\end{abstract}

\maketitle
\section{\label{sec:intro}Introduction}

The accurate modeling of alloys with first-principles methods is a difficult computational effort in the materials science community \cite{VANDEWALLE2009266}. The main issue arises from the usually unknown stoichiometry and local atomic ordering of the alloyed structures. One way to combat this is to manually create supercells with a specific stoichiometry and local atomic ordering determined empirically or intuitively \cite{GaSe-optical,C7CP06750J,KANLI201913}. Another, more systematic option is to create a large set of alloyed structures (of varying compositions and stoichiometries) and determine the energy hull diagram of the alloy series from first-principles coupled with statistical methods \cite{VANDEWALLE2009266,wines-nitride,bowing}. Popular methods usually involve the Special Quasirandom (SQS) \cite{PhysRevLett.65.353} generation of several structures, evaluation of energies from first-principles, and then using the first-principles energies as a training set for statistical methods such as the Cluster Expansion \cite{VANDEWALLE2009266} formalism.

Constructing an accurate energy hull diagram is essential for predicting the energetic stability and favorability of alloys. The accuracy of an energy hull diagram depends critically on the choice of the method used to generate total energies. Density functional theory (DFT) has been used for this purpose, thanks to its balance between accuracy and computational scaling on modern computer clusters. On the other hand, diffusion Monte Carlo (DMC) \cite{RevModPhys.73.33,Needs_2009} is another ground state method which can provide total energy differences with significantly better accuracy than DFT, with minimal functional and basis set dependence \cite{doi:10.1063/5.0023223,PhysRevMaterials.5.024002,PhysRevX.9.011018,bilayer-phos,annaberdiyev2021cohesion}. DMC has a very similar computational scaling to DFT, $N^{3-4}$, though with nearly a thousand times larger prefactor, but its embarrassingly parallel algorithm can utilize parallel computation architectures using larger number of nodes. Recently, DMC has been used to calculate chemically accurate total energies and electronic properties for a wide variety of two-dimensional (2D) and bulk structures, demonstrating the routine applicability of this method \cite{PhysRevMaterials.5.024002,ataca_qmc,PhysRevLett.115.115501,bilayer-phos,doi:10.1063/1.5026120,PhysRevB.95.081301,PhysRevB.96.119902,PhysRevB.96.075431,PhysRevB.101.205115,PhysRevMaterials.3.124414,PhysRevX.9.011018,PhysRevMaterials.2.085801,Luo_2016,C6CP02067D,doi:10.1063/1.4919242,PhysRevMaterials.1.073603,PhysRevMaterials.2.075001,PhysRevB.98.155130,phosphors,LiNiO2,PhysRevB.93.094111,annaberdiyev2021cohesion,doi:10.1063/5.0022814,bennett2021origin}. In this work, building on our previous DMC study of 2D GaSe \cite{doi:10.1063/5.0023223}, we propose to use DMC to reliably calculate the energetics of 2D GaS$_{x}$Se$_{1-x}$ as a case study. This will enable accurate simulations of alloy properties for experimental characterization and validation. 


DMC calculations require multiple steps to obtain the total energy of a structure and the stochastic nature of the method introduces uncertainty in the observables, in contrast to the deterministic nature of DFT. Constructing an adequate energy hull diagram requires the high-throughput calculation of the energy of several structures. Similarly to the development of the decade old field of high throughput DFT \cite{JAIN20112295}, researchers have attempted to bridge the gap between high throughput computing and Quantum Monte Carlo (QMC) and increase the computational efficiency of the QMC workflow \cite{Zen1724,ht-qmc-saritas}. For example, Saritas et al. \cite{ht-qmc-saritas} developed a high throughput procedure to determine the  DMC formation energy of certain materials directly from the ICSD database. In this work, we propose a new high throughput workflow to determine the alloy formation energy of 2D GaS$_{x}$Se$_{1-x}$ with near chemical accuracy. We went on to test various wavefunction optimization methods, which included recycling Jastrow parameters (which we call \textit{Jastrow sharing}) among various alloy configurations to reduce the overall computational cost and localization errors.

In addition to 2D GaS$_{x}$Se$_{1-x}$ being a suitable case study that builds off our previous QMC work \cite{doi:10.1063/5.0023223}, it has been reported that alloying is a promising route to control the properties of several monolayer or few layer materials \cite{ingaalloy,KANLI201913,TiS,C7CP06750J,Li-ads,photoresp,ersan-alloy,ataca-alloy}  (including post transition metal chalcogenides, or PTMCs \cite{Hosseini_Almadvari_2020,wines-nitride,PhysRevB.102.075414,doi:10.1002/adma.201601184,bowing}), which makes the study of 2D alloys with accurate QMC methods useful for theorists and guiding experimentalists in synthesis and characterization. For example, it has experimentally been reported that when GaSe nanostructures are alloyed with Te, the material undergoes a hexagonal to monoclinic transition and there exists an instability region where the phases compete and two different band gaps can be found at the same composition \cite{bowing}. Janus monolayers \cite{PhysRevApplied.13.064008,PhysRevB.102.075414} (a class of two-dimensional 2D alloyed structures) have also gained attention due to their applications for Schottky contacts, as demonstrated theoretically with Janus GaSSe on top of graphene \cite{PhysRevB.102.075414}. In addition, PTMCs such as GaSe and GaS can be engineered for specific applications by chemical functionalization \cite{o-func,C5CP00397K,pyradine}, creating heterostructures \cite{PhysRevB.102.075414,Lie1501882,C8CP03740J,gate-mos2,JAPPOR2017109}, and applying strain \cite{PhysRevApplied.11.024012,doi:10.1142/S0217984915500499,C5NR08692B}. PTMCs such as GaSe have also been reported to be suitable substrate material for other 2D structures, as well \cite{D0CP00357C,PhysRevMaterials.2.104002}.

 In section \ref{sec:methods} we outline our DFT and QMC approaches and convergence criteria. In section \ref{sec:cluster} we present our DFT results in order to initially screen 2D GaS$_{x}$Se$_{1-x}$ alloys. In section \ref{sec:jastrow} we describe our \textit{Jastrow sharing} methodology and present a detailed analysis of this methodology at the variational Monte Carlo (VMC) and DMC level. In section \ref{sec:finitesize} we present DMC results using various supercell sizes for alloys and compare extrapolated results using different wavefunction optimization methods. In section \ref{sec:cost} we give an analysis of the computational cost savings of our approach. Finally, we provide concluding remarks and future perspectives in section \ref{sec:conclusion}.   
 
\section{\label{sec:methods}Computational Methods and Theory}
Benchmarking DFT calculations were performed using the VASP code with projector augmented wave (PAW) potentials \cite{PhysRevB.54.11169,PhysRevB.59.1758}. For these VASP benchmarking calculations, the Perdew-Burke-Ernzerhof (PBE)\cite{PhysRevLett.77.3865} and strongly constrained and appropriately normed (SCAN)\cite{PhysRevLett.115.036402} meta-GGA functionals were used. In addition, the PBE+D2 \cite{doi:10.1002/jcc.20495} and PBE+D3 \cite{doi:10.1063/1.3382344} methods of Grimme and the SCAN+rvv10 \cite{PhysRevX.6.041005} functionals were used to investigate vdW effects at the DFT level. At least 20 \AA\space of vacuum was given between periodic layers of GaS$_x$Se$_{1-x}$ in the $c$-direction, a kinetic energy cutoff of 350 eV was used, and a 6x6x1 reciprocal grid was used for the primitive cells of the alloys and the number of k-points were scaled appropriately with supercell size. 

DMC calculations use wavefunctions from other ab-initio methods, in our case DFT-PBE, as input (trial wavefunction) to obtain an equilibrium that is computationally tractable. For DFT calculations within the QMC workflow, the Quantum Espresso (QE) \cite{Giannozzi_2009} code was used. For Ga, S, and Se we used norm-conserving Burkatzki-Filippi-Dolg (BFD) \cite{doi:10.1063/1.2741534,doi:10.1063/1.2987872} pseudopotentials. DMC calculations require norm-conserving pseudopotentials and BFD pseudopotentials have been thoroughly tested in DMC \cite{doi:10.1063/1.4991414, doi:10.1063/5.0023223}. For details of pseudopotential testing, conversion and validation, refer to the discussion and Table S1 in the Supplementary Information (SI). For the pseudopotentials, we used a kinetic energy cutoff of 160 Ry (see Fig. S1), which gives a convergence of less than 1 meV at the DFT level. We used the same supercell reciprocal twist grid in QE as our benchmarking VASP calculations which is tested in Fig. S2.

After the trial wavefunction is generated using DFT, VMC and DMC \cite{RevModPhys.73.33,Needs_2009} calculations were carried out using the QMCPACK \cite{Kim_2018,doi:10.1063/5.0004860} code. VMC calculations are the intermediate steps between the DMC and DFT calculation, where the single determinant DFT wavefunction is converted into a many-body wavefunction, through the Jastrow parameters \cite{PhysRev.34.1293,PhysRev.98.1479}. Jastrow parameters help model the electron correlation and reduce the uncertainty in the DMC calculations \cite{PhysRevLett.94.150201,doi:10.1063/1.460849}. The automated DFT-VMC-DMC workflows were generated using the Nexus \cite{nexus} software suite. 
Up to three-body Jastrow \cite{PhysRevB.70.235119} correlation functions were included. The linear method \cite{PhysRevLett.98.110201} was used to minimize the variance and energy respectively of the VMC total energies . The cost function of the variance optimization is 100$\%$ variance minimization and the cost function of the energy optimization is split as 95$\%$ energy minimization and 5$\%$ variance minimization, which has been shown to reduce uncertainty for DMC results \cite{PhysRevLett.94.150201}. The explicit details of how we modified this optimization procedure for our high-throughput QMC workflow will be given in section \ref{sec:jastrow}.

Being a real-space wavefunction method, DMC calculations need to be performed at increasing supercell sizes to eliminate finite-size errors. We used supercell sizes up to 72 atoms and extrapolated to the infinite cell size. In order to smooth the image interactions at each supercell used in the extrapolation, we used the {\fontfamily{qcr}\selectfont{optimal\_tilematrix}} function in Nexus, which allows constructing appropriate supercells with the largest Wigner-Seitz (WS) radius for a given size. We used Jackknife fitting to obtain a linear fit of DMC data and extrapolate to the infinite-size limit for total ground state energy of each alloyed structure. In addition, the locality approximation \cite{doi:10.1063/1.460849} was used to evaluate the nonlocal pseudopotentials in DMC. A timestep of 0.01 Ha$^{-1}$ (convergences tests are shown in Fig. S3) was used for all DMC simulations. 

To initially create and screen these GaS$_x$Se$_{1-x}$ alloyed structures with varying concentrations, the Special Quasirandom Structure (SQS) method \cite{PhysRevLett.65.353} was employed as implemented in the ATAT code \cite{VANDEWALLE2009266}. The alloy formation energy in units of eV/formula unit (otherwise known as the mixing enthalpy) is defined as:

\begin{equation}
E_\textrm{{form}}=E_{{\textrm{GaS}_{1-x}\textrm{Se}_{x}}}-(1-x)E_\textrm{{GaS}}-xE_\textrm{{GaSe}}
\label{formformula}
\end{equation} 
where $E_{{\textrm{GaS}_{1-x}\textrm{Se}_{x}}}$ is the energy of a GaS$_x$Se$_{1-x}$ cell, $E_\textrm{{GaS}}$ is the energy of a GaS cell, and $E_\textrm{{GaSe}}$ is the energy of a GaSe cell, all per formula unit (f.u.). After the SQS generation of random alloys, first-principles calculations are done to calculate the formation energy of each structure. These energies are then used to obtain fitted energies via the Cluster Expansion formalism \cite{VANDEWALLE2009266}. Further details and convergence criteria can be found in the Supplementary Information (SI).

\section{\label{sec:results}Results and Discussion}

\subsection{\label{sec:cluster} DFT Calculations}

\begin{figure*}
\begin{center}
\includegraphics[width=14.5cm]{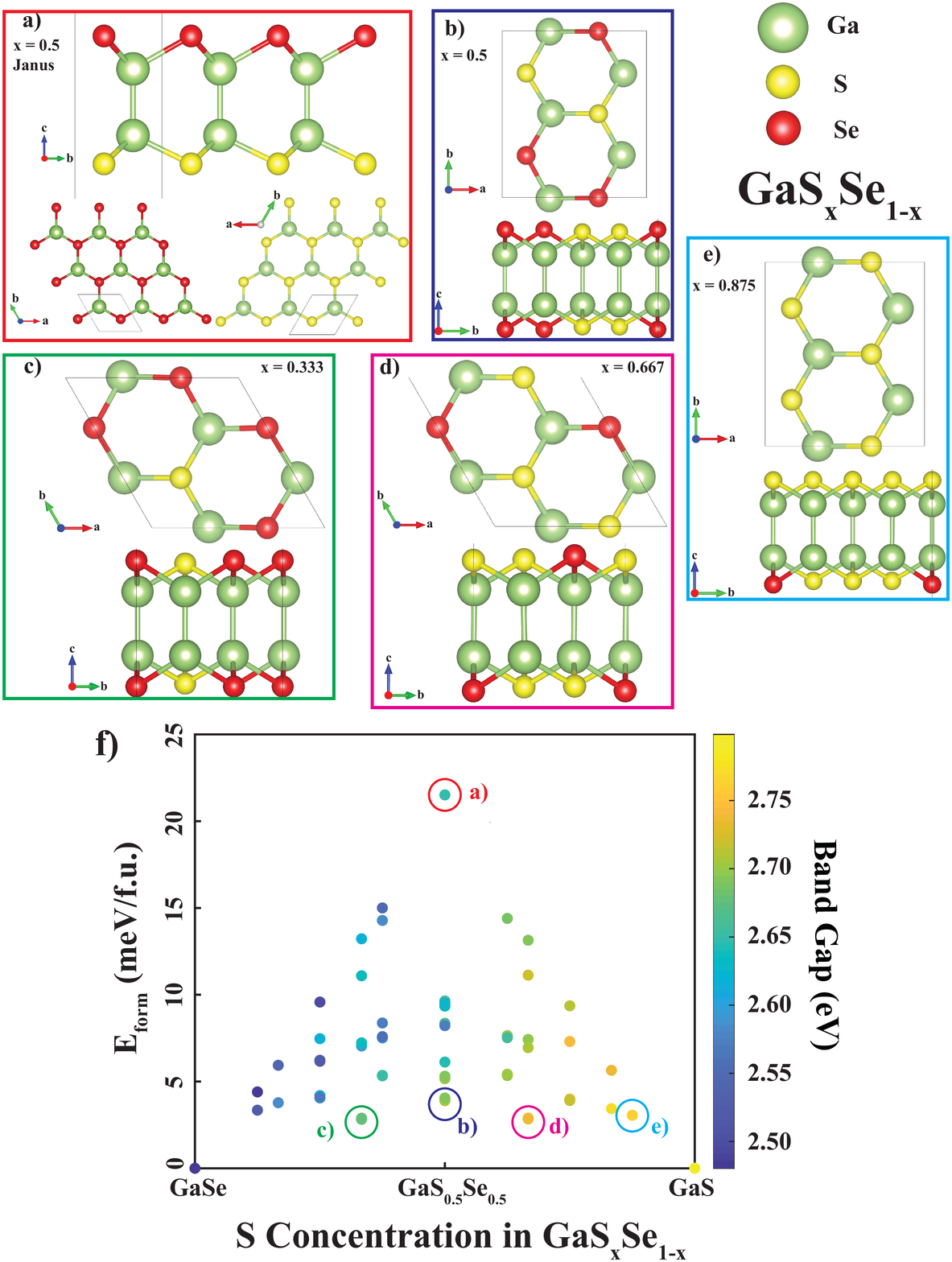}
\caption{The atomic structure (top and side view) of selected SQS GaS$_x$Se$_{1-x}$ alloys from the energy hull diagram: a) Janus GaS$_{0.5}$Se$_{0.5}$, b) GaS$_{0.5}$Se$_{0.5}$ (lowest E$_{\textrm{form}}$), c) GaS$_{0.333}$Se$_{0.667}$ (lowest E$_{\textrm{form}}$), d) GaS$_{0.667}$Se$_{0.333}$ (lowest E$_{\textrm{form}}$), e) GaS$_{0.875}$Se$_{0.125}$ (smallest Wigner-Seitz radius) and f) the energy hull diagram obtained using DFT (SCAN+rvv10, PAW potentials), where the structures depicted (a - e) are circled with the corresponding color outline and the band gap is given in the color axis for each structure.}
\label{geofig}
\end{center}
\end{figure*}

In contrast to a previous DFT study where the 2D GaS$_{x}$Se$_{1-x}$ alloys were constructed manually \cite{GaSe-optical}, we used the SQS method to generate alloys and DFT to initially optimize the geometries. Lattice parameters of 2D GaS and GaSe have been determined experimentally \cite{GaSe-optical}, therefore accurate determination of the geometry is an important step for the respective 2D alloys, which have unknown experimental lattice constants. Due to this, we performed geometry and lattice optimization calculations with PBE, PBE-D2, PBE-D3, SCAN and SCAN+rvv10 DFT functionals to benchmark. This process also allows us to construct the energy hull diagrams using each functional (see Fig. \ref{geofig} and Fig. S4). We find that all the tested functionals are in good agreement with each other in their formation energies, but lattice parameters differ slightly for each structure. However, we find that the SCAN+rvv10 functional gives the overall best agreement with the experimental lattice parameters of 2D GaS and GaSe \cite{GaSe-optical,GaSe-nature-synthesis} (see Table S2 in SI). This is expected since it has been previously reported that SCAN yields lattice constants and optimal geometry in closer agreement to experiment for several 2D materials \cite{buda,doi:10.1063/5.0023223}. Therefore, we use the geometries from the SCAN+rvv10 functional as a starting point for DMC calculations. To ensure that the optimal geometry of these structures are closest to exact and do not bias the energy, we isotropically scaled the lattice (starting from the geometry and lattice constants obtained from SCAN+rvv10) and confirmed what the location of the minimum energy was with respect to lattice constant for the equation of state (see Fig S5-9). The benchmarking of lattice parameters with different functionals are done using VASP because the plane wave cutoff of PAW potentials are more than an order of magnitude smaller than BFD pseudopotentials.

As seen in Fig. \ref{geofig} f), all the monolayer GaS$_{x}$Se$_{1-x}$ structures have positive formation energy, which implies that these structures are potentially not energetically favorable at zero temperature and cannot form spontaneously in vacuum (metastable). In addition, Janus GaS$_{0.5}$Se$_{0.5}$ (where S and Se atoms are on the opposing faces of the 2D layer), has the highest formation energy, which means that this structure is the least energetically favorable in free-standing form. This is a similar finding to Ersan \textit{et al.} \cite{PhysRevApplied.13.064008}, where it was found that monolayer Janus PtXY (X = S, Se, Te) is the least favorable alloyed structure in the 2D PtXY alloy series. Even though these reported Janus structures are thermodynamically stable (as a result of phonon and molecular dynamics simulations) in vacuum \cite{PhysRevApplied.13.064008}, they can perhaps be synthesized on a substrate material \cite{janus-graphene,janus-tmd}. In free-standing form, molecular dynamics simulations showed that certain Janus structures can spontaneously curl \cite{C8CP02011F}. Figure S4 and \ref{geofig} f) also reflects the band gap tunability (color axis) of GaS$_{x}$Se$_{1-x}$ monolayers at the DFT level, opening the door to various electronic applications. Following the initial geometric optimization and construction of the energy hull diagram with DFT, we selected certain structures to further run QMC on to test our \textit{Jastrow sharing} procedure which we will discuss in the next subsection.

\subsection{\label{sec:jastrow}Jastrow Parameter Sharing}

In attempt to reduce the number of wavefunction optimization simulations required to perform DMC for all the 2D GaS$_{x}$Se$_{1-x}$ structures involved in the Cluster Expansion formalism, we propose a systematic method we call \textit{Jastrow sharing}. Wavefunction optimization using Jastrow parameters is key to reducing variance and the localization error of subsequent DMC simulations \cite{PhysRevB.93.094111,doi:10.1063/1.4991414,doi:10.1063/1.4986951}. Jastrow parameters are typically optimized using the cheaper VMC calculations, but these costs can be important considering the large number of materials simulated in the Cluster Expansion formalism. Computational cost analysis will further be discussed in Section \ref{sec:cost}. Our \textit{Jastrow sharing} approach involves a judicious selection of optimized Jastrow parameters from a single compound, and reusing it in all the rest of the materials studied. In this section we discuss the various methods we used to test and validate this method for GaS$_{x}$Se$_{1-x}$ and the limitations of this method with respect to other material classes.  

In our method, we use a supercell that has the smallest Wigner-Seitz (WS) radius to obtain the optimized Jastrow parameters that will later be used in all DMC calculations. The maximum Jastrow cutoff radius possible in a structure is equal to its WS radius, which is equal to the radius of the largest inscribing sphere that can fit into the simulation cell. Even though the Jastrow cutoff can be optimized, this increases the computational scaling of the VMC optimization \cite{PhysRevB.70.235119}, hence it is more practical to use the maximum Jastrow cutoff possible for a periodic system (WS radius), and update each Jastrow parameter at a fixed cost. Therefore, it is imperative that the smallest WS radius among all the structures is determined in advance in order to reuse the same set of Jastrow parameters for all the structures. For three-body Jastrows, a cutoff radius of 2 \AA\space (on the order of GaS$_{x}$Se$_{1-x}$ bond lengths) was used. Due to the fact that three-body Jastrows (energy optimization) usually capture shorter-range correlations, we made this arbitrary choice to minimize computational cost while attempting to capture such interactions. The full workflow of our procedure is given in Fig. \ref{flowchart}.

We performed DMC calculations on four structures: x = 0.333 (12 atom cell, Fig. \ref{geofig} c)), x = 0.5 (16 atom cell, Fig. \ref{geofig} b)) and x = 0.667 (12 atom cell, Fig. \ref{geofig} d)) and the Janus GaS$_{0.5}$Se$_{0.5}$ structure (highest formation energy, 16 atom cell, Fig. \ref{geofig} a)). Except for the Janus structure, all these compounds lie on the bottom of the DFT hull diagram. However, this selection can be made on a finite threshold above the DFT hull or more ambitious plans can involve all the SQS generated structures. The reason we decided to investigate materials with low and high formation energy is to observe if QMC can correctly capture these energy differences and to demonstrate that our method can work for the entire energy hull.

\begin{table}[]
\begin{tabular}{c|c|c|c|c}

\toprule
{ \textbf{DMC-J3}} & { \textbf{Janus}} & { \textbf{x = 0.5}} & { \textbf{x = 0.333}} & { \textbf{x = 0.667}} \\
\toprule
Janus Jastrow                          & 0                                     & -6(6)                                  & 2(6)                                    & 3(10)                                    \\
\hline
\hline
x =   0.5 Jastrow                      & 0(4)                                 & 0                                       & 4(7)                                     & 15(11)                                    \\
\hline
\hline
x =   0.333 Jastrow                    & -1(4)                                & -2(4)                                   & 0                                        & 17(10)                                     \\
\hline
\hline
x =   0.667 Jastrow                    & -6(5)                                & -10(6)                                  & -1(6)                                    & 0                                         \\
\hline
\hline
x =   0.875 Jastrow                    & -2(5)                                & -15(7)                                  & -2(5)                                    & 9(9)                                    \\
\toprule
{ \textbf{DMC-J2}} & { \textbf{Janus}} & { \textbf{x = 0.5}} & {\textbf{x = 0.333}} & { \textbf{x = 0.667}} \\
\toprule
Janus Jastrow                          & 0                                    & -3(7)                                  & -9(7)                                    & -6(7)                                    \\
\hline
x =   0.5 Jastrow                      & 3(4)                                 & 0                                       & 5(7)                                     & -9(8)                                    \\
\hline
x =   0.333 Jastrow                    & -7(5)                                & 4(9)                                   & 0                                         & 6(9)                                     \\
\hline
x =   0.667 Jastrow                    & -9(7)                                & -2(8)                                  & -2(6)                                    & 0                                         \\
\hline
x =   0.875 Jastrow                    & -8(4)                                & -5(8)                                  & -2(6)                                    & -3(7)      \\
\toprule                                   
\end{tabular}
\caption{The differences (in meV/f.u.) between total energies (DMC) calculated with \textit{self-Jastrows} and \textit{shared-Jastrows} including up to two-body and up to three-body terms with the associated error bars in parenthesis. The rows represent which Jastrows are used and the columns represent each alloyed structure for which the DMC energies are calculated.}
\label{sharing}
\end{table}

Among the unit cells of the alloyed structures we selected to study, x = 0.875 has the smallest WS radius (see Fig. \ref{geofig} e)). Therefore, according to our procedure, we would use GaS$_{0.875}$Se$_{0.125}$ to generate the Jastrow parameters to be used in the remainder of the calculations. However, in order to understand the sensitivity of our method to this decision, we make a benchmark study where we optimize Jastrow parameters of all the structures separately as it is typically done in DMC calculations. For testing, a 16 atom cell was used for x = 0.5, Janus, 0.875 and a 12 atom cell was used for x = 0.333, 667. To be able to use all 5 of these sets of Jastrow parameters across all the structures we study, we still use the smallest WS radius as the cutoff parameter for two-body Jastrow parameters. The smallest WS radius we have in this case is 3.12 \AA. This is longer than 2 \AA, hence no modifications are required for 3-body Jastrows. The testing of our \textit{Jastrow sharing} procedure is tabulated in Table \ref{sharing}, where we show the energy difference (in meV/f.u.) between the total energy of each structure using its own optimized Jastrow parameters (we will define this as a \textit{self-Jastrows}) and the total energy of each structure using the optimized Jastrow parameters of a different structure (we will define these as \textit{shared-Jastrows}), with the associated error. In order to accept the \textit{Jastrow-sharing} procedure as valid, we expect the localization error obtained with \textit{self-Jastrows} and \textit{shared-Jastrows} to be comparable and the total energy difference (per f.u.) between \textit{self-Jastrows} and \textit{shared-Jastrows} to be as minimal as possible. 

In Table \ref{sharing}, the columns represent the different selected structures and the rows represent the various optimized Jastrow parameters. The upper quadrant contains the energy difference at the DMC level using up to three-body Jastrows and the lower quadrant contains the energy difference at the DMC level using up to two-body Jastrows. A more detailed table with VMC energies, including two- and three-body Jastrows are given in Table S3 in the SI. From the data in Table \ref{sharing}, we observe that at the DMC level, the total energy differences between \textit{shared-Jastrows} and \textit{self-Jastrows} are overall smaller, with differences ranging from 0 - 17 meV/f.u. We also observe that the energy differences calculated with DMC are nearly identical whether or not two-body or three-body interactions are included in the Jastrow factor due to the fact that we don't have enough resolution to differentiate between the results. The lack of significant variations in the DMC energies here also suggest that the localization errors of these pseudopotentials are very small or negligible since the quality of the Jastrow parameters, with regards to the Jastrow cutoff, do not affect the DMC calculations.  

\begin{figure*}
\begin{center}
\includegraphics[width=14.5cm]{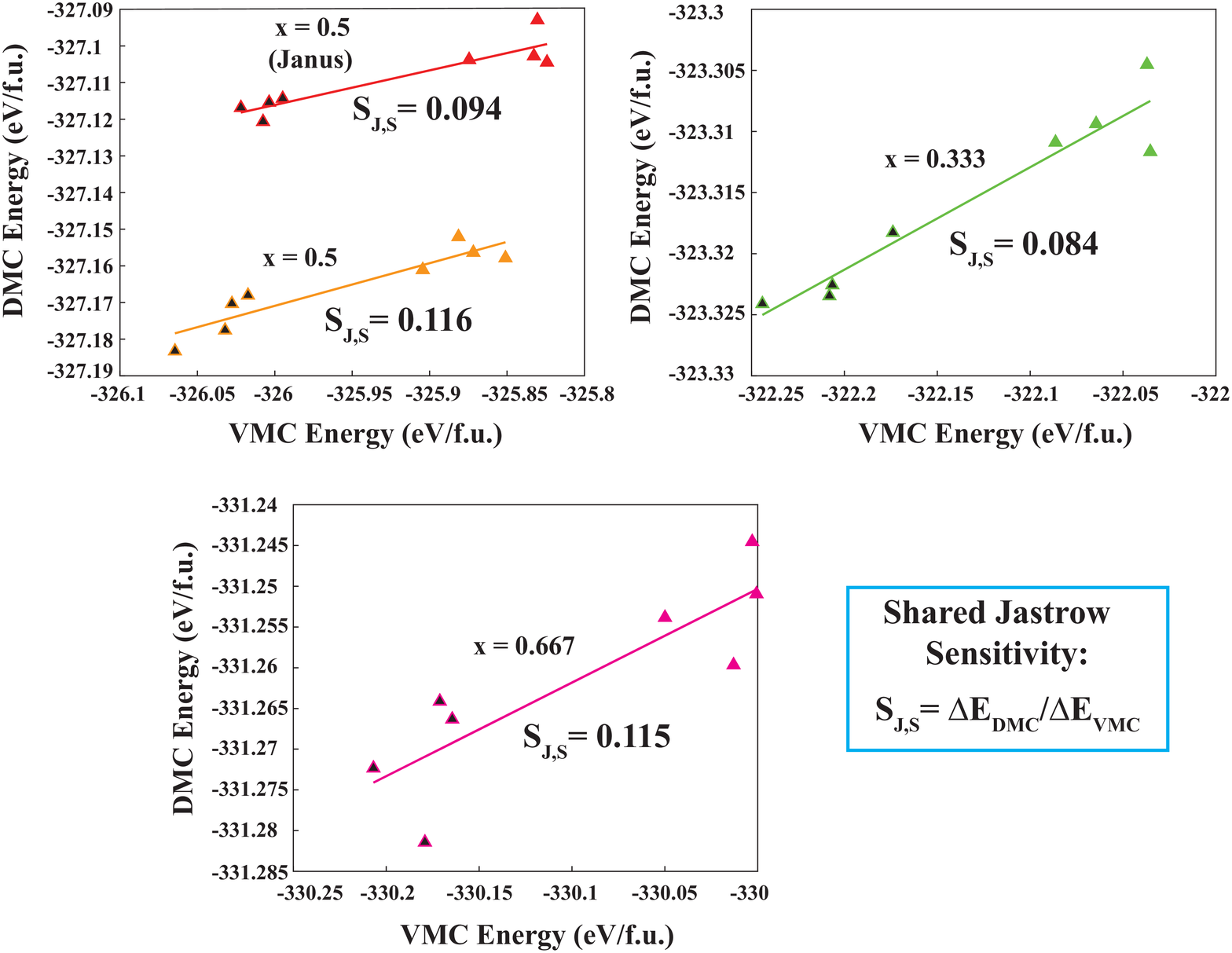}
\caption{The total DMC energy (eV/f.u.) as a function of total VMC energy (eV/f.u.) for the Janus, x = 0.5, x = 0.333, and x = 0.667 structures. The formula for \textit{shared-Jastrow sensitivity} (per f.u.), which is defined as the change in DMC energy per change in VMC energy (slope of the graph), is given in each inset. The data points with a black center represent J3 results while the other points represent J2 results. Due to their non linearity, the data points using the Janus Jastrow were excluded (see SI). }
\label{sensitivity}
\end{center}
\end{figure*}

In Figure \ref{sensitivity}, we additionally show that DMC-J3 total energies are always lower than the DMC-J2 energies. This has been observed for a variety of systems \cite{doi:10.1063/1.4986951}. However, in a practical sense, the variation in a set of Jastrow parameters (J2 or J3) is more important than the separation between the DMC energies calculated with J2 and J3 Jastrow parameters. This is because often the user makes one selection regarding the number of Jastrow parameters and applies it throughout for all similar materials studied in the same calculation set, such as polymorphs or alloys. Ideally, the quality of the Jastrow parameters should only change the effort required to get the target uncertainty in the DMC total energies. However, as observed in here, we often see that the quality of the Jastrow parameters also change the degree of localization error \cite{PhysRevB.93.094111,doi:10.1063/1.4991414,doi:10.1063/1.4986951} which can lead to inconsistencies when energy differences are calculated between two structures.

In order to quantify the amount of localization error introduced as a function of the quality of the Jastrow parameter, we can utilize a quantity called Jastrow sensitivity. The Jastrow sensitivity is defined as the amount of energy decrease in the DMC energy per unit decrease in the VMC energy and is a property of the pseudopotential used in the calculation \cite{doi:10.1063/1.4986951}. Ideally, the Jastrow sensitivity should be small for a given pseudopotential for all of the relevant valence enviroments (i.e. different phases of a given material).  We infer that a low enough Jastrow sensitivity could be a reason why we were successfully able to transfer different Jastrow parameters across the 2D GaS$_{x}$Se$_{1-x}$ series. In theory, if the sensitivity of the pseudopotentials is low for the material class of study, our \textit{Jastrow sharing} procedure should be valid. 

To further quantify Jastrow sensitivity in our work, we define a new quantity called \textit{shared-Jastrow sensitivity} (S$_{J,S}$), which is defined as the DMC energy per unit decrease in the VMC energy calculated with \textit{self-Jastrows} and \textit{shared-Jastrows}. In order to calculate this for each structure, we plotted the DMC total energies vs. the VMC total energies (per f.u.) and obtained the slope for the data (using J2 and J3 parameters, from Table \ref{sharing}), which is depicted in Fig. \ref{sensitivity}. Due to the fact that VMC energies obtained with the Janus Jastrow parameters have a large deviation attributed to cutoff radius, we excluded these points from the linear fitting. A more detailed discussion related to the Jastrow parameters of the Janus structure can be found in the SI. From Fig. \ref{sensitivity} we find the \textit{shared-Jastrow sensitivity} (per f.u.) to be 0.094 (0.047 per atom) for the Janus structure, 0.116 (0.058 per atom) for the x = 0.5 structure, 0.084 (0.042 per atom) for the x = 0.333 structure, and 0.115 (0.058 per atom) for the x = 0.667 structure. 

These sensitivities are comparable in value to the lower bounds on the Jastrow sensitivities reported by Krogel and Kent\cite{doi:10.1063/1.4986951} (0.05 for Ce 4+ using the locality approximation) and Dzubak, Krogel, and Reboredo\cite{doi:10.1063/1.4991414} (0.07, 0.05, 0.04 for Mn, Fe, Co respectively). In addition to our calculated \textit{shared-Jastrow sensitivity}, we calculated the standard Jastrow sensitivity of the individual Ga, S and Se atoms to further prove transferability across different atomic systems. We obtained this by calculating the DMC energy with no Jastrow parameters and the DMC energy using up to J2 and J3 Jastrows, then fitting a line to the three data points. We obtain a Jastrow sensitivity of 0.07, 0.14 and 0.04 for Ga, S and Se respectively. As a result of these sensitivity calculations, we can quantify that the Jastrows can be transferable between various stoichiometries in the GaS$_x$Se$_{1-x}$ system. If one wished to implement this \textit{Jastrow-sharing} procedure for another system, calculating the \textit{shared-Jastrow sensitivity} using VMC and DMC energies from \textit{self-Jastrows} and \textit{shared-Jastrows} and the standard Jastrow sensitivity of the respective atoms and comparing to tabulated results in literature could be a viable method to check if \textit{Jastrow-sharing} is feasible.

\subsection{\label{sec:finitesize}Finite-size Effects}

\begin{table}[h]
\begin{tabular}{c|c|c|c}
\toprule
 & \multicolumn{3}{c}{\textbf{Cohesive Energy (eV/f.u.)}} \\
\toprule
Structure & \begin{tabular}[c]{@{}c@{}}DFT\\ (PBE)\end{tabular} & \begin{tabular}[c]{@{}c@{}}DMC\\ self-Jastrow\end{tabular} & \begin{tabular}[c]{@{}c@{}}DMC\\ WS-Jastrow\end{tabular} \\
\toprule
GaSe                                                         & -7.135                                              &      -7.028(3)\cite{doi:10.1063/5.0023223}                                                 & -                                                              \\
\hline
GaS                                                        & -7.789                                              &     -7.659(4)                                                  & -                                                              \\
\hline
Janus                                                       & -7.360                                              &   -7.231(6)                                                    &     -7.240(6)                                                 \\
\hline
x = 0.5                                                     & -7.456                                              &    -7.35(1)                                                    &      -7.35(2)                                                 \\
\hline
x = 0.333                                                   & -7.347                                              &    -7.259(6)                                                  &    -7.258(6)                                                  \\
\hline
x = 0.667                                                    & -7.564                                              &        -7.461(6)                                               &    -7.464(7)                                                   \\

\toprule
\toprule
 & \multicolumn{3}{c}{\textbf{Formation Energy (eV/f.u.)}} \\
\toprule
Structure & \begin{tabular}[c]{@{}c@{}}DFT\\ (PBE)\end{tabular} & \begin{tabular}[c]{@{}c@{}}DMC\\ self-Jastrow\end{tabular} & \begin{tabular}[c]{@{}c@{}}DMC\\ WS-Jastrow\end{tabular} \\
\toprule
Janus                                                      & 0.1019                                              &      0.106(4)                                                  &    0.096(4)                                                    \\
\hline
x = 0.5                                                     & 0.0059                                              & 0.01(1)                                                        & 0.00(2)                                                        \\
\hline
x = 0.333                                                    & 0.0057                                              &     0.006(4)                                                  &     0.007(4)                                                  \\
\hline
x = 0.667                                                    & 0.0064                                              &       0.013(3)                                                &    0.010(5)                                                   \\

\toprule
\end{tabular}
\caption{The calculated cohesive energy (top half) and alloy formation energy (bottom half) all in eV/f.u. obtained from DFT (PBE, BFD pseudopotentials) and DMC (using \textit{self-Jastrows} and \textit{WS-Jastrows}) for the Janus, x = 0.5, x = 0.333, and x = 0.667 structures.}
\label{formtable}
\end{table}

\begin{figure*}
\begin{center}
\includegraphics[width=14.5cm]{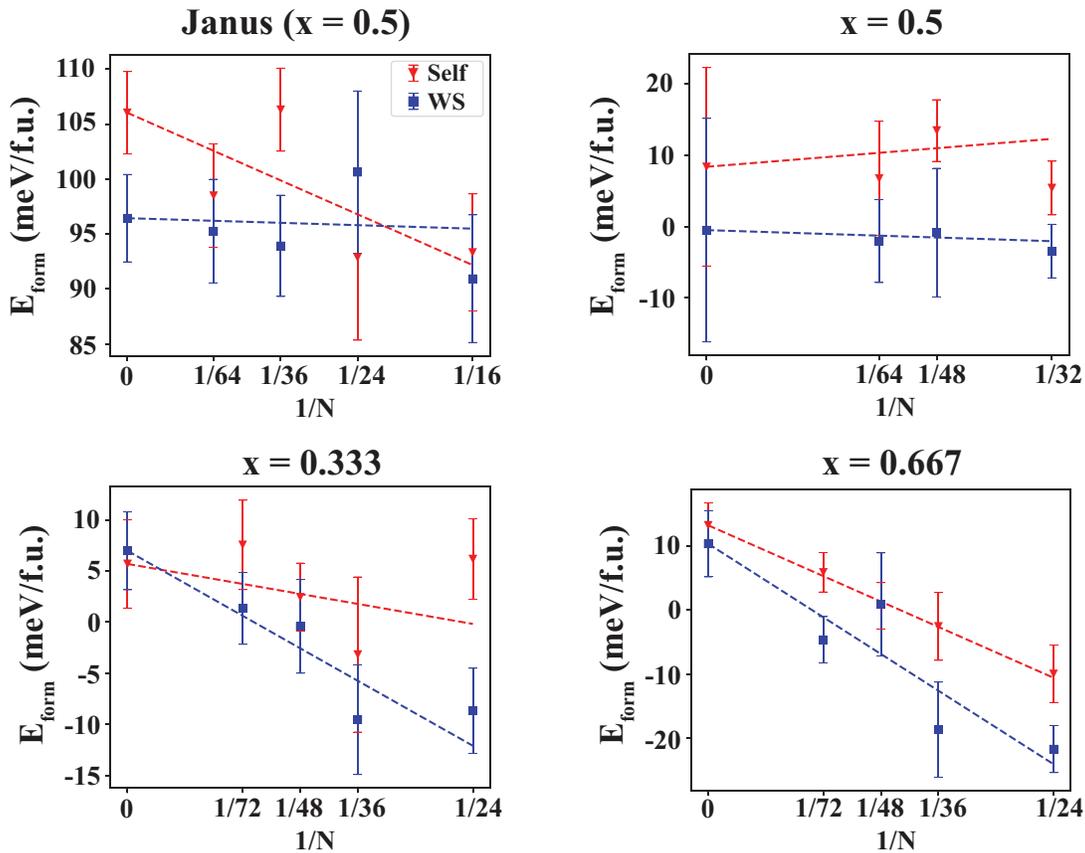}
\caption{The DMC calculated formation energies as a function of inverse number of atoms (N) in the supercell, extrapolated to the thermodynamic limit (N$\rightarrow\infty$) for the Janus, x = 0.5, x = 0.333, and x = 0.667 structures. Red represents DMC energies calculated with \textit{self-Jastrows} while blue represents DMC energies calculated with \textit{WS-Jastrows}.}
\label{extrap}
\end{center}
\end{figure*}

\begin{figure}
\begin{center}
\includegraphics[width=7.5cm]{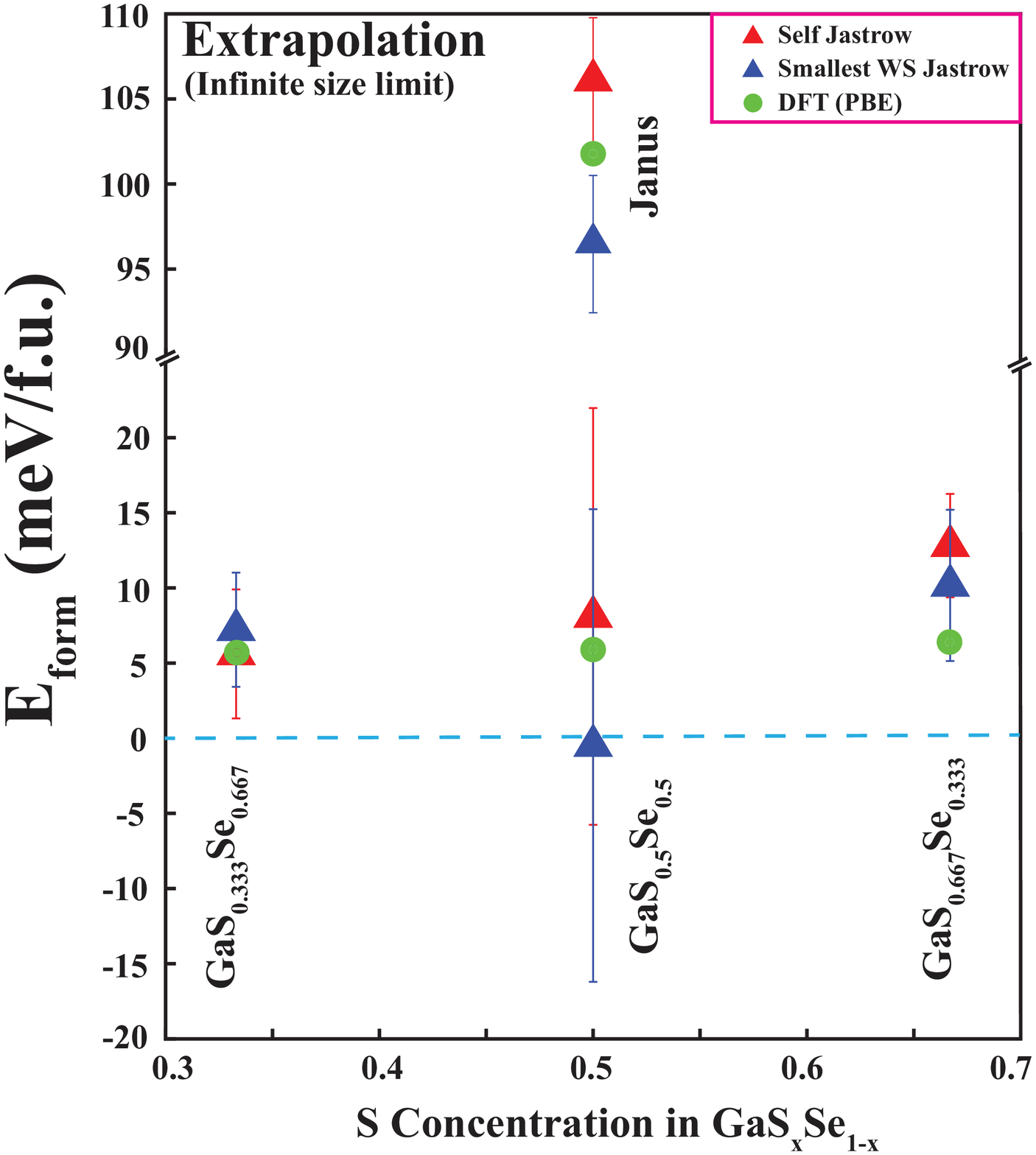}
\caption{ The partially constructed energy hull diagram for GaS$_x$Se$_{1-x}$ calculated with DFT (PBE) in green and DMC (using \textit{self-Jastrows} in red and \textit{WS-Jastrows} in blue), with associated error bars.}
\label{supercell}
\end{center}
\end{figure}

To construct our DMC energy hull diagram and eliminate finite-size effects, we extrapolate all structures to the thermodynamic limit. Therefore, our DMC calculations are performed at different supercell sizes and then extrapolated to the infinite sized limit as the energy scales as 1/N, where N is the number of particles. In addition to this, we use a converged reciprocal grid obeying periodic boundary conditions for each supercell, which is known as twist averaging. For an additional benchmark to assess the validity of our approach, we calculated the DMC cohesive energy (per f.u.) of each structure by subtracting the appropriate single atom energies of Ga, Se and S (see SI for single atom calculation details) from the extrapolated total energy. We calculated the alloy formation energy ($E_\textrm{{form}}$) at the DMC level using equation \ref{formformula} (see Table \ref{formtable}). The total energy of the GaSe and GaS supercells with the exact same number of atoms as the specific alloy supercell was subtracted from the total energy of said alloy. This alloy formation energy was calculated at multiple supercell sizes and extrapolated to the thermodynamic limit (see Figure \ref{extrap}). 

For the DMC extrapolated cohesive energies and formation energies, \textit{self-Jastrows} were used for each respective supercell. In addition, the optimized Jastrow parameters of the alloy with the smallest WS radius (we will refer to this as \textit{WS-Jastrow}, with parameters taken from the 16 atom GaS$_{0.875}$Se$_{0.125}$ cell) were used to calculate the total DMC energy for each supercell size. A comparison of the DMC results (for extrapolated cohesive energies and alloy formation energies) using \textit{self-Jastrows} and \textit{WS-Jastrows} and comparison to DFT is found in Table \ref{formtable}. In addition, the partial energy hull diagram for the selected structures using DFT and DMC (with \textit{self-Jastrows} and \textit{WS-Jastrows}) is depicted in Figure \ref{supercell}. From the linear extrapolated results (Fig. \ref{extrap}) depicted in Fig. \ref{supercell} and tabulated in Table \ref{formtable}, we clearly see the energies calculated (and then extrapolated) using \textit{self-Jastrows} and \textit{WS-Jastrows} are nearly indistinguishable. 

These extrapolated formation energies demonstrate that an adequate energy hull diagram can be constructed from DMC using just one set of Jastrow parameters from the smallest WS cell, in contrast to optimizing the Jastrows for every structure at every supercell size. Additionally, using a many-body approach such as QMC for alloys can provide near-chemically accurate confirmation of stability and other important electronic and energetic properties. 2D GaS$_x$Se$_{1-x}$ was chosen because it is a convenient system to test this methodology on. This is convenient because it builds on our previous 2D GaSe \cite{doi:10.1063/5.0023223} work and since there is relatively good agreement between DFT functionals for the formation energy hull diagram (Fig. S4), it allows us to benchmark the resolution of our DMC extrapolated (to meV resolution) energies with reliable DFT energies. 

\begin{figure}
\begin{center}
\includegraphics[width=5.5cm]{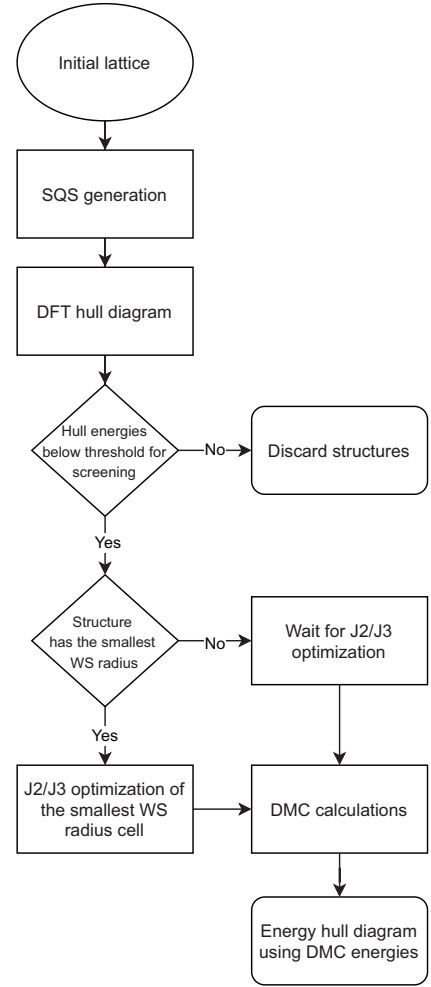}
\caption{The full high throughput workflow proposed in this work to obtain the energetics of an alloy system with QMC methods.  }
\label{flowchart}
\end{center}
\end{figure}

A full schematic of this workflow is given in Fig. \ref{flowchart}. This workflow starts with an input of an initial lattice, followed by the SQS generation of random alloys and then the construction of the energy hull diagram with DFT. After analyzing the DFT hull diagram, the user can choose to run QMC for structures below a certain energy threshold or discard them. From the SQS structures, the user must determine which has the smallest WS radius, and then perform J2 and J3 Jastrow optimization for such. These J2 and J3 parameters are then reused for DMC calculations of the other selected structures. From this, the DMC energy diagram can be constructed. A more detailed analysis of the cost savings of this workflow is given in the following section.

\subsection{\label{sec:cost}Computational Cost Analysis}

\begin{table}[]
\begin{tabular}{c|c|c|c}

\multicolumn{4}{c}{\textbf{VMC Computational Time}}\\
\hline
 & { \textbf{1 Supercell}} & { \textbf{Extrapolation}} & \textbf{Energy Hull}\\
\hline
Self Jastrow                          & $755 \pm 180$        & $7,124 \pm 1,816$   & $24,043 \pm 5,881$                                                  \\
\hline
WS Jastrow                      & $134 \pm 40$    & $134 \pm 40$  & $134 \pm 40$       \\
\hline
\multicolumn{4}{c}{\textbf{}}\\

\multicolumn{4}{c}{\textbf{DMC Computational Time}}\\ 
\hline
 & {\textbf{1 Supercell}} & {\textbf{Extrapolation}} & {\textbf{Energy Hull}} \\
\hline
Self Jastrow                          & $2,143 \pm 1,071$      & $32,899 \pm 16,450$   & $93,066 \pm 46,533$                                                             \\
\hline
WS Jastrow                      & $2,109 \pm 1,054$   & $33,259 \pm 16,630$ &   $93,170 \pm 46,616$       

\end{tabular}
\caption{A measure of the computational cost (time in seconds) using Self Jastrows and WS Jastrows for 1 supercell, one extrapolated result, and the entire energy hull diagram for VMC (upper half) and DMC (lower half) simulations. The quantity in the $\pm$ gives the upper and lower bound on cost (discarding unnecessary QMC steps after target error is reached).}
\label{costtablevmc}
\end{table}

We claim that reusing a fixed single set of Jastrow paramaters across all structures involved in creating a DMC energy hull diagram is a valid and computationally efficient approach. Therefore, we compare the efficiency of our procedure to another more conventional way of performing these DMC calculations, where the Jastrow parameters are optimized separately for each structure including the different sized supercells used in the finite size extrapolations. Otherwise, all the settings in the DMC and VMC calculations are uniform across all the calculations, including the number of cores and number of walkers used. In this section we analyze the computational cost associated with \textit{Jastrow sharing}. We can see from our results in Table \ref{sharing} that the VMC and DMC energies are converging to the appropriate ground states with comparable localization error using the same VMC and DMC parameters on the same number of nodes/cores, which demonstrates that \textit{Jastrow sharing} for GaS$_x$Se$_{1-x}$ is a valid approach. Therefore, we report the computational time for these two procedures. 

Before proceeding any further, we provide details of our computational environment. All of the VMC and DMC simulations are conducted on 8 nodes of our local cluster. Each node is connected via OmniPath networking and contains 2 Intel 20-Core Xeon Gold 6138 CPUs clocked at 2GHz and 384 GB of 2666MHz DDR4 memory. QMCPack version 3.9.0 is compiled using OpenMPI version 3.0.1 and Intel Fortran, C and C++  (version 19.0.4.243) compiler.

To quantitatively investigate the computational cost savings of \textit{Jastrow sharing}, we report the  computational time of VMC and DMC (total time including every twist) in Table \ref{costtablevmc}. We report the computational time obtained with parameters we used in our QMC simulations (upper bound) and calculate a lower bound, which is obtained by removing the number of extra unnecessary steps after a target uncertainty was reached ($\sim$10 meV/f.u. for VMC, $\sim$5 meV/f.u. for DMC). We used the 36 atom supercell of GaS$_{0.333}$Se$_{0.667}$ as an example of a single supercell (second column of Table \ref{costtablevmc}). As seen from Table \ref{costtablevmc} (second column), the VMC time associated with optimizing the Jastrow parameters of the smallest WS structure is much lower due to the fact that the WS structure only has 16 atoms (while the other supercell has 36 atoms).

To report total cost savings for all of the calculations performed in this paper, we give the VMC and DMC time needed to achieve the extrapolated results (using x = 0.333 as an example, see third column of Table \ref{costtablevmc}). We also report the computational time to achieve the extrapolated results for the four considered structures (Janus, x = 0.5, x = 0.333, and x = 0.667) to get an estimate of the cost savings for constructing the energy hull diagram (see Fig. \ref{supercell}).  Since the DMC extrapolated results using the \textit{WS-Jastrow} only requires one VMC optimization simulation (extra VMC calculations are avoided), the VMC time for the extrapolation is the same as the single supercell (third column of Table \ref{costtablevmc}). The substantial savings of constructing the hull diagram (at the VMC level) is indicated in the fourth column of Table \ref{costtablevmc}, where the VMC computational time of using \textit{WS-Jastrows} is two orders of magnitude smaller than using \textit{self-Jastrows}. 

From Table \ref{costtablevmc}, we observe that using either the \textit{self-Jastrows} or the \textit{WS-Jastrow} does not change the DMC computational time significantly, which means that \textit{Jastrow-sharing} does not increase cost at the DMC level in our case. Specifically, each set of Jastrow parameters are of comparable quality and do not increase the uncertainty or amount of DMC steps needed to reach the target uncertainty for each calculation. It is clear from this cost analysis that by sharing Jastrow parameters, it is possible to achieve substantial savings by reducing the VMC computational time. The cost of VMC calculations is often disregarded when compared to DMC cost, but in our work we observe that when \textit{self-Jastrows} are used, VMC accounts for roughly 1/4 (on average) of the total cost. Therefore by using the \textit{WS-Jastrow} approach, the high-throughput QMC calculations of alloy systems can be significantly less costly.

\section{\label{sec:conclusion}Conclusion}

We have outlined a high-throughput procedure to calculate the energy hull diagram for an alloy system using QMC methods. This involves optimizing the Jastrow parameters of the alloy with the smallest WS radius, and using these optimized parameters for all of the other subsequent structures. We tested the validity of this \textit{Jastrow-sharing} procedure on monolayer GaS$_{x}$Se$_{1-x}$ by swapping the Jastrow parameters of structures with different stoichiometries and tabulated the Jastrow sensitivities. Finally, we quantified the substantial computational cost savings obtained from avoiding extra VMC optimization simulations. We hope this method can be implemented for other systems where the Jastrow sensitivities of the subsequent pseudopotentials are low, which can aid in future accurate and high-throughput studies for alloys.

\section*{Supplementary Material}

See the supplementary information for additional details and convergence tests of DFT and DMC calculations, convergence criteria for the SQS and Cluster Expansion, DFT and DMC details about the optimal geometries of each structure, testing of \textit{Jastrow-sharing} at the VMC level, and details about finite-size extrapolation.

\begin{acknowledgments}
This work was supported by the National Science Foundation through the Division of Materials Research under NSF DMR-1726213.
\end{acknowledgments}

\section*{Data Availability}

The data that support the findings of this study are available from the corresponding author upon reasonable request.

\section*{References}
\nocite{*}
\bibliography{Draft}

\end{document}


\section{DFT and DMC convergence testing and additional details}

\begin{figure}
\begin{center}
\includegraphics[width=8.5cm]{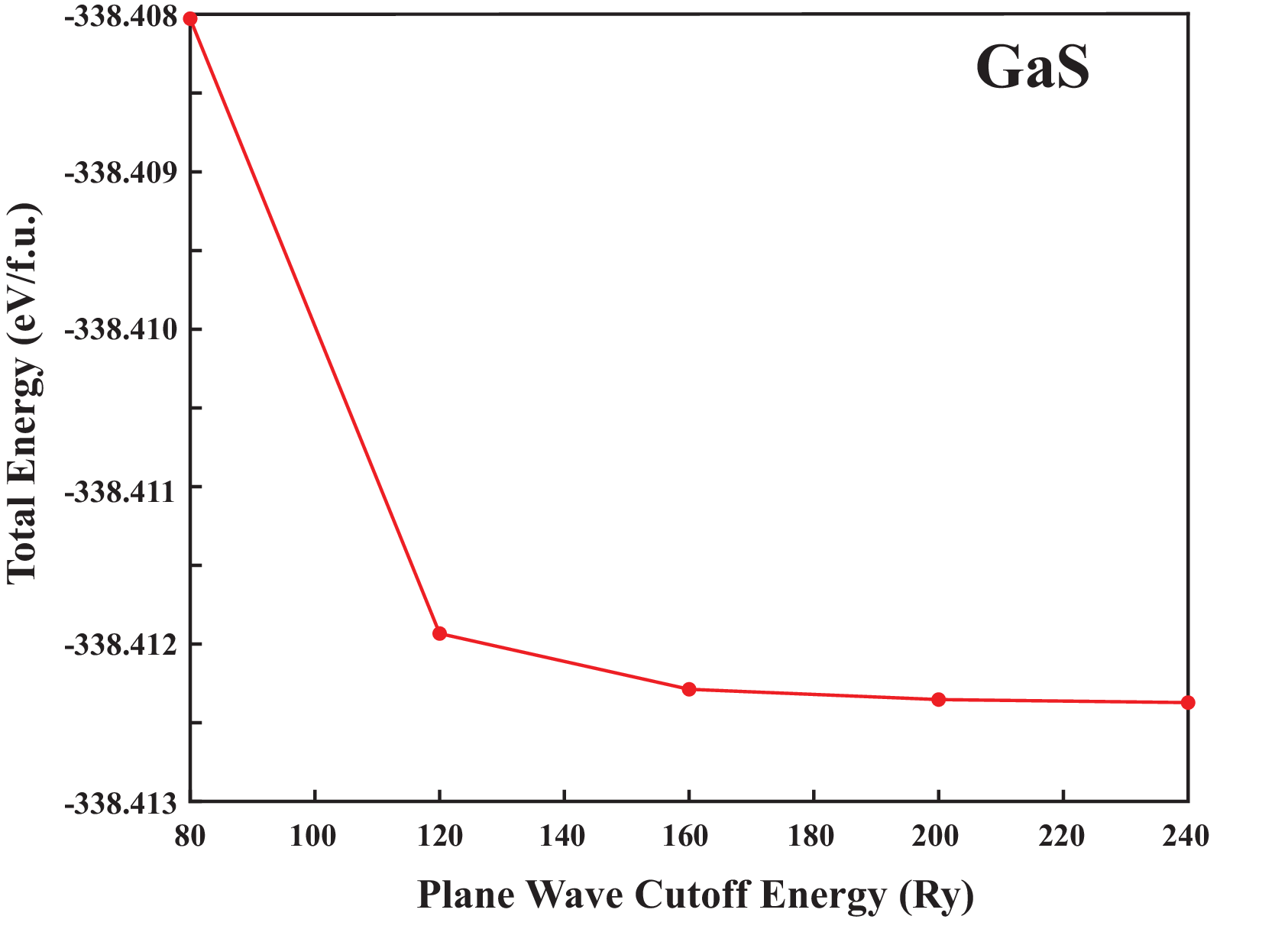}
\caption{The total energy per formula unit of 2D GaS as a function of plane wave cutoff energy for the BFD pseudopotentials converted to plane wave format (calculated with PBE). The results show a converged value of 160 Ry.}
\label{encut}
\end{center}
\end{figure}

\begin{figure}
\begin{center}
\includegraphics[width=8.5cm]{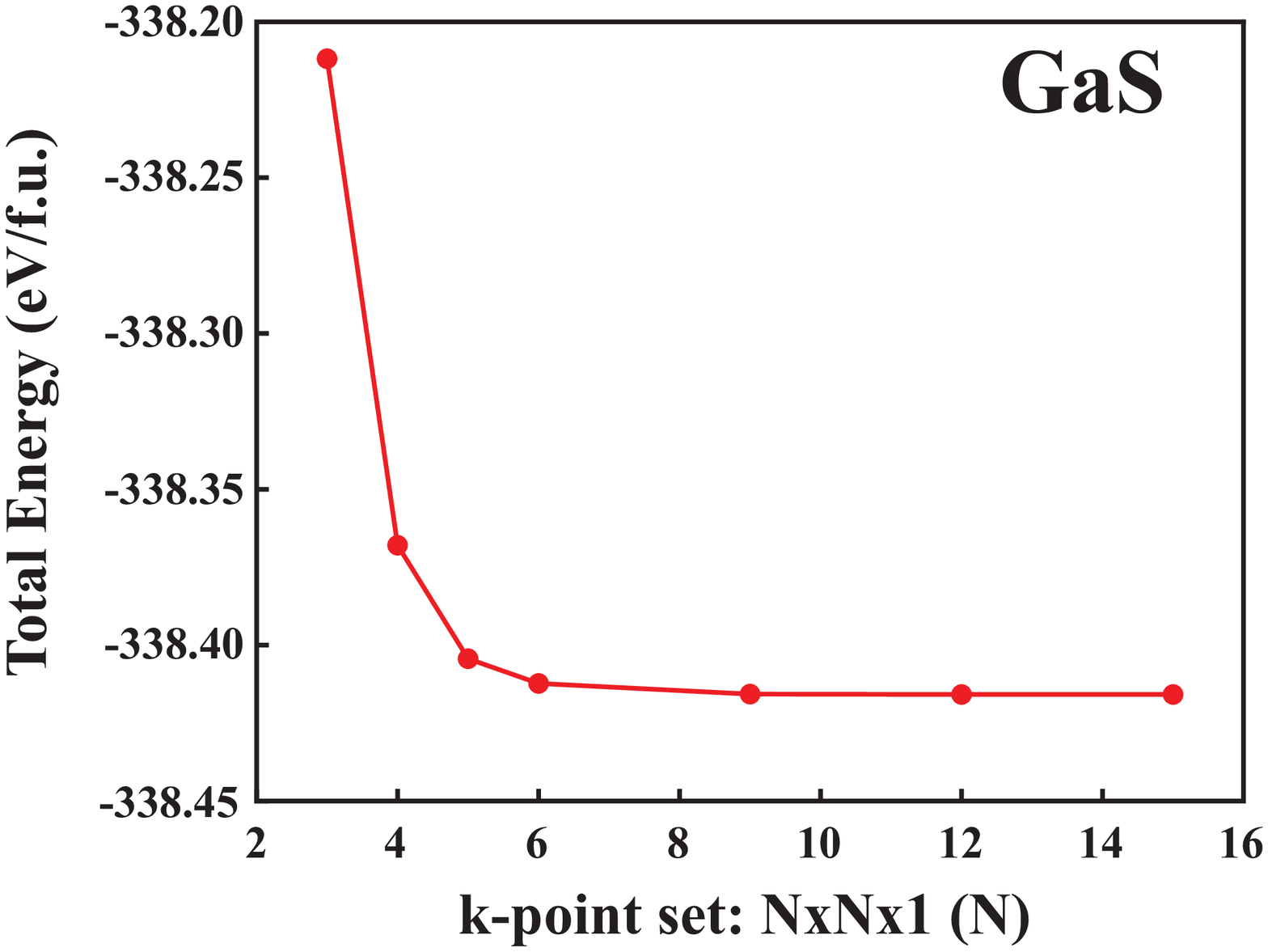}
\caption{The total energy per formula unit of 2D GaS as a function of k-point grid for the BFD pseudopotentials converted to plane wave format (calculated with PBE). The results show a converged k-point grid of 6x6x1.}
\label{kpts}
\end{center}
\end{figure}

\begin{table}

\caption{\label{tab:tableSI}%
To quantify the conversion of pseudopotentials, the calculated ionization potential and electron affinity for Ga, Se and S atoms calculated with the converted plane wave format BFD potentials (using PBE) are given. The experimental values are also given as a reference.}

\begin{tabular}{c|c|c}
\hline
\hline
 Method   & Ionization Potential (eV)  & Electron Affinity (eV)  \\
 \hline
 \hline
PBE (BFD)  & Ga: 5.89\cite{doi:10.1063/5.0023223}, Se: 9.44\cite{doi:10.1063/5.0023223}, S: 10.14  & Ga: 0.32\cite{doi:10.1063/5.0023223}, Se: 2.15\cite{doi:10.1063/5.0023223}, S: 2.20  \\
Exp     & Ga: 6.00\cite{lide_2008}, Se: 9.75\cite{lide_2008}, S: 10.36\cite{lide_2008} & Ga: 0.30\cite{PhysRevA.100.052512}, Se: 2.02\cite{PhysRevA.85.015401}, S: 2.08\cite{s-electronaffinity}  \\

\hline
\hline
\end{tabular}
\label{table:bulk}

\end{table}

In order to convert these Gaussian potentials to plane wave format, we used the ppconvert \cite{Kim_2018} tool implemented in QMCPACK. We converted these Gaussian potentials to plane wave format and validated this conversion by calculating the ionization potential (IP) and electron affinity (EA) at the DFT level. We used the p local channel for these BFD potentials. For our DFT benchmarking calculations in VASP, we used standard PBE PAW (as opposed to the hard or soft versions) potentials from version 5.4.4. 

In the ATAT code, what is known as the Cluster Expansion (CE) formalism is performed \cite{VANDEWALLE2009266}. In CE, the first-principles formation energies from the SQS structures are used as a training set to calculate fitted energies. CE can be used to validate the training set size of the SQS alloy and even predict the energetics of structures outside of the training set. When the fitted and calculated energies match within a convergence criteria known as the cross validation score, which is designed to estimate the error made in predicting the fitted energy of a structure and is analogous to the root mean square error, we know that we have considered enough random structures in our training set and can verify our predictions. In our CE calculations, we allowed the minimum cross validation score to be lower than 25 meV (thermal energy fluctuations at room temperature). \cite{VANDEWALLE2009266} We achieved a smaller cross validation value after 60 simulations of alloys at varying concentrations.

\begin{figure}
\begin{center}
\includegraphics[width=14cm]{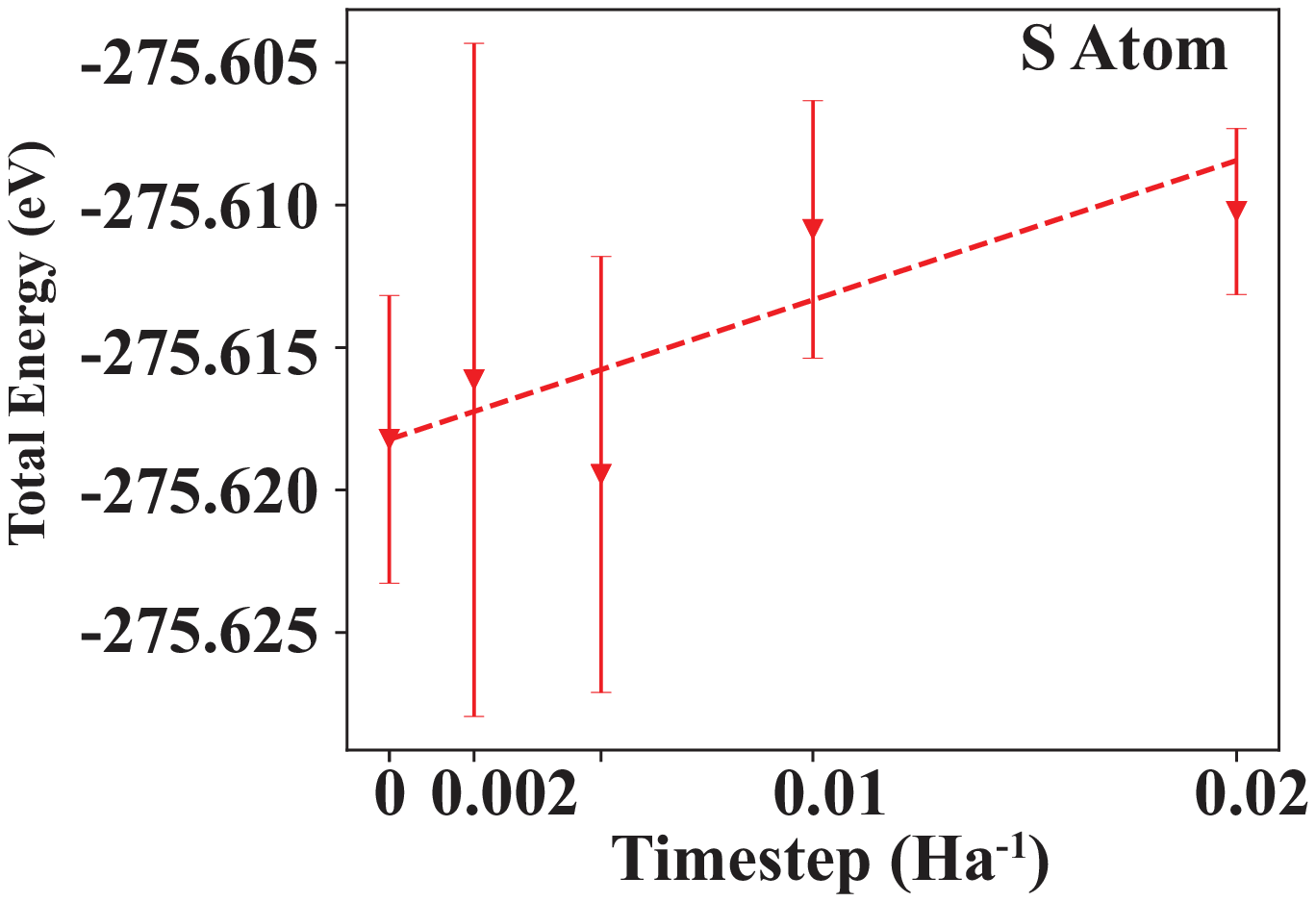}
\caption{The timestep convergence test for the total energy of a single S atom. A timestep of 0.01 Ha$^{-1}$ was used in the calculation of cohesive energy.}
\label{stime}
\end{center}
\end{figure}

\begin{figure}
\begin{center}
\includegraphics[width=16.5cm]{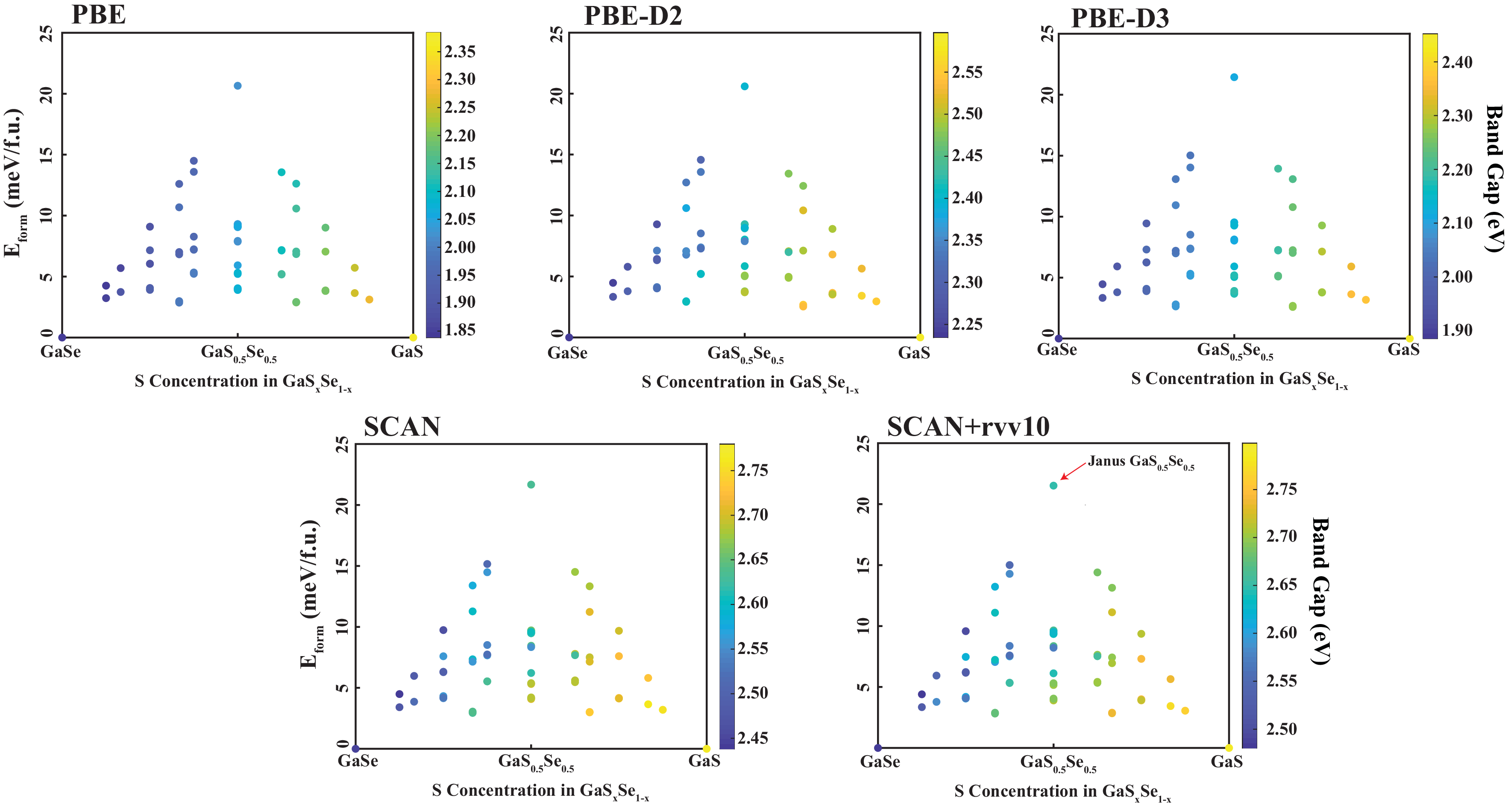}
\caption{The formation energy per formula unit as a function of S concentration of 2D GaS$_{x}$Se$_{1-x}$ alloys created by the SQS method. Each data point represents a different alloyed structure and the structure is optimized and energy is calculated with the PBE, PBE-D2, PBE-D3, SCAN and SCAN+rvv10 functionals. The color axis gives the calculated band gap of each structure.}
\label{CE1}
\end{center}
\end{figure}

\begin{figure}
\begin{center}
\includegraphics[width=10.5cm]{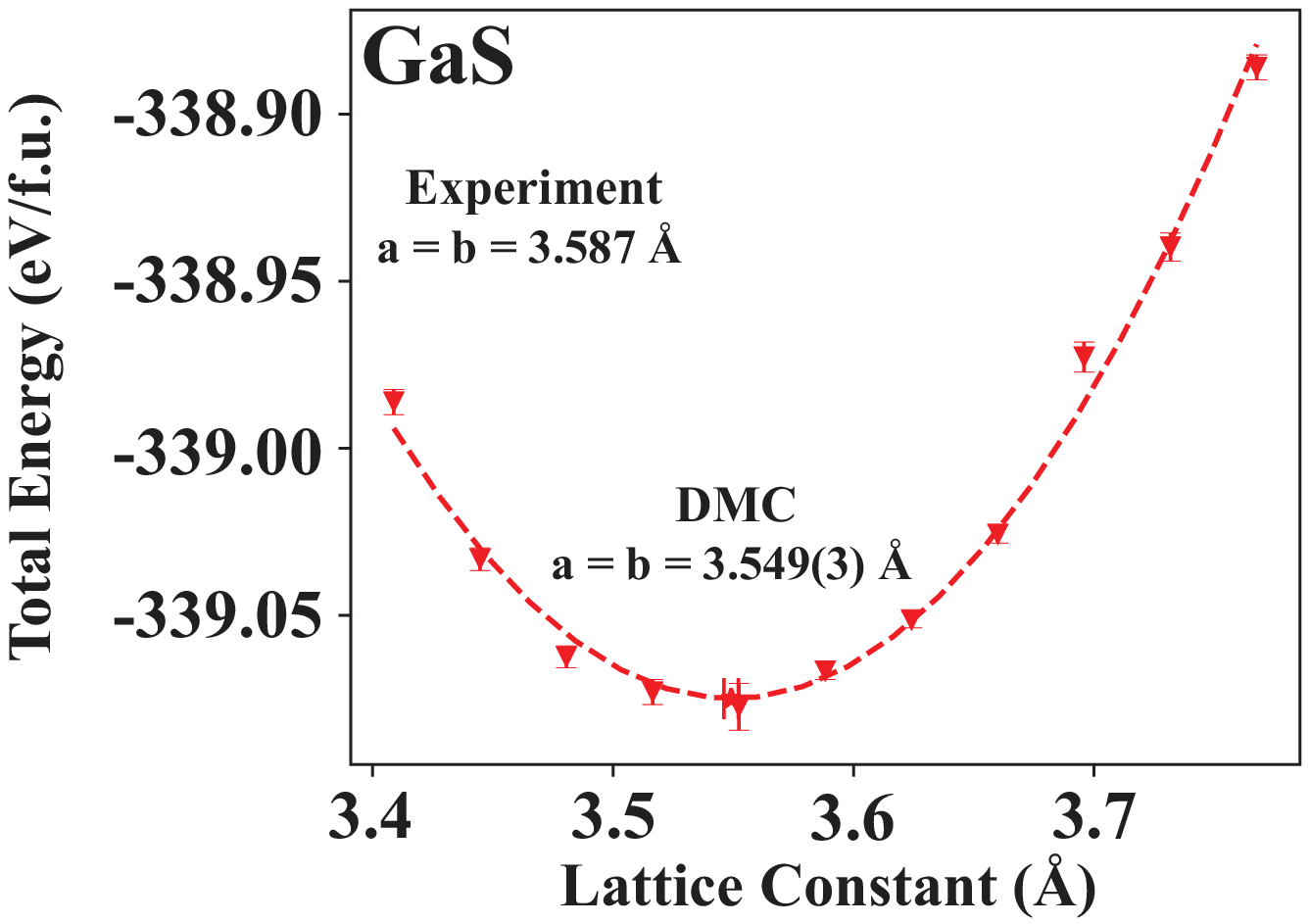}
\caption{The total energy per formula unit of 2D hexagonal GaS (16 atom cell) calculated with DMC using starting wavefunctions from PBE versus the isotropically scaled lattice constant (in the $x$ and $y$ direction). The dotted line represents the fitted curve and the minimum point and respective error bar is marked on the curve (with the value given in the figure inset). The experimental lattice constant from reference \cite{GaSe-optical} is also given in the inset for comparison. }
\label{gas-geo}
\end{center}
\end{figure}

\begin{figure}
\begin{center}
\includegraphics[width=10.5cm]{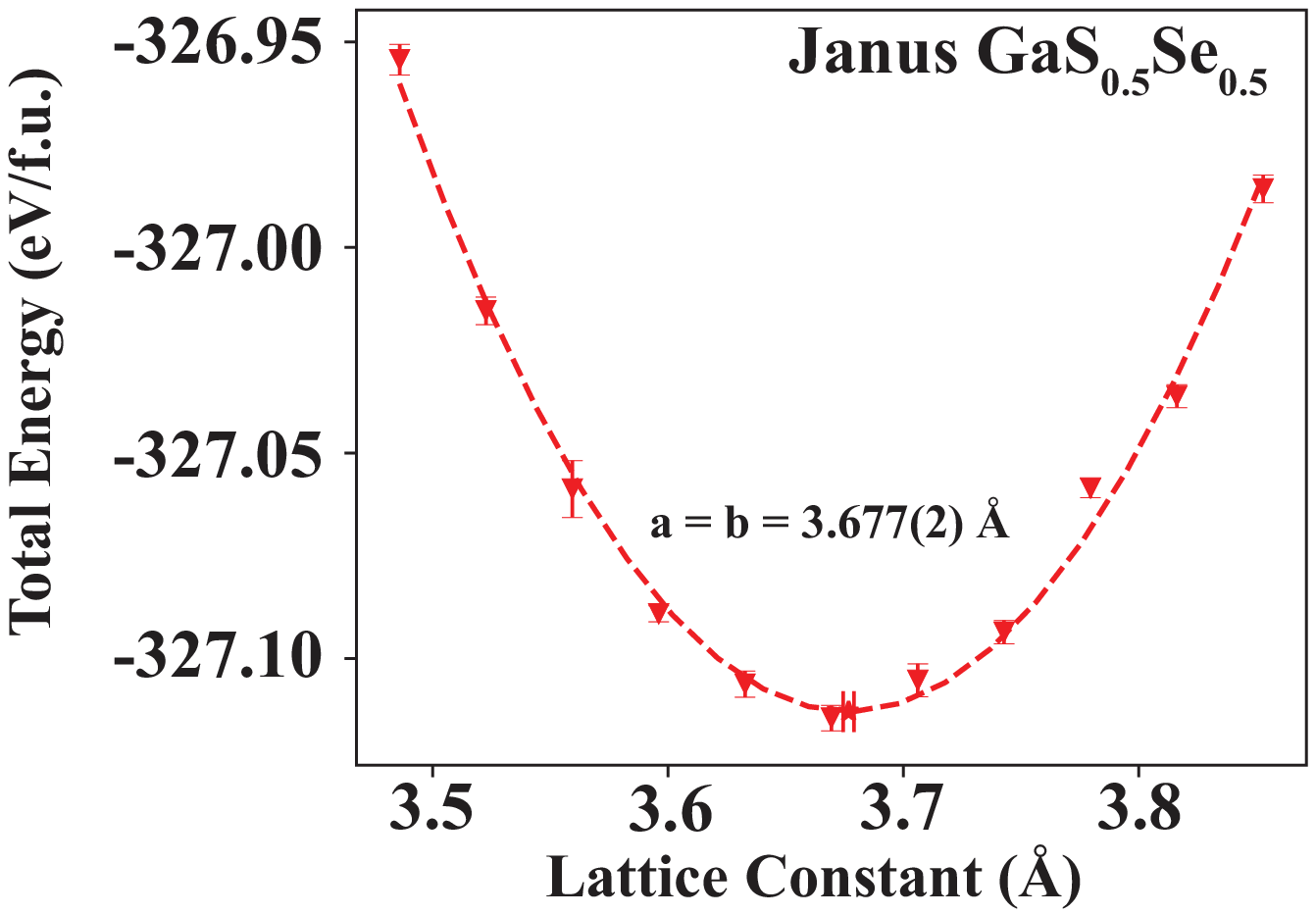}
\caption{The total energy per formula unit of 2D hexagonal Janus GaS$_{0.5}$Se$_{0.5}$ (16 atom cell) calculated with DMC using starting wavefunctions from PBE versus the isotropically scaled lattice constant (in the $x$ and $y$ direction). The dotted line represents the fitted curve and the minimum point and respective error bar is marked on the curve (with the value given in the figure inset). Geometric structure is depicted in Fig. 1 a). }
\label{janus-geo}
\end{center}
\end{figure}

\begin{figure}
\begin{center}
\includegraphics[width=10.5cm]{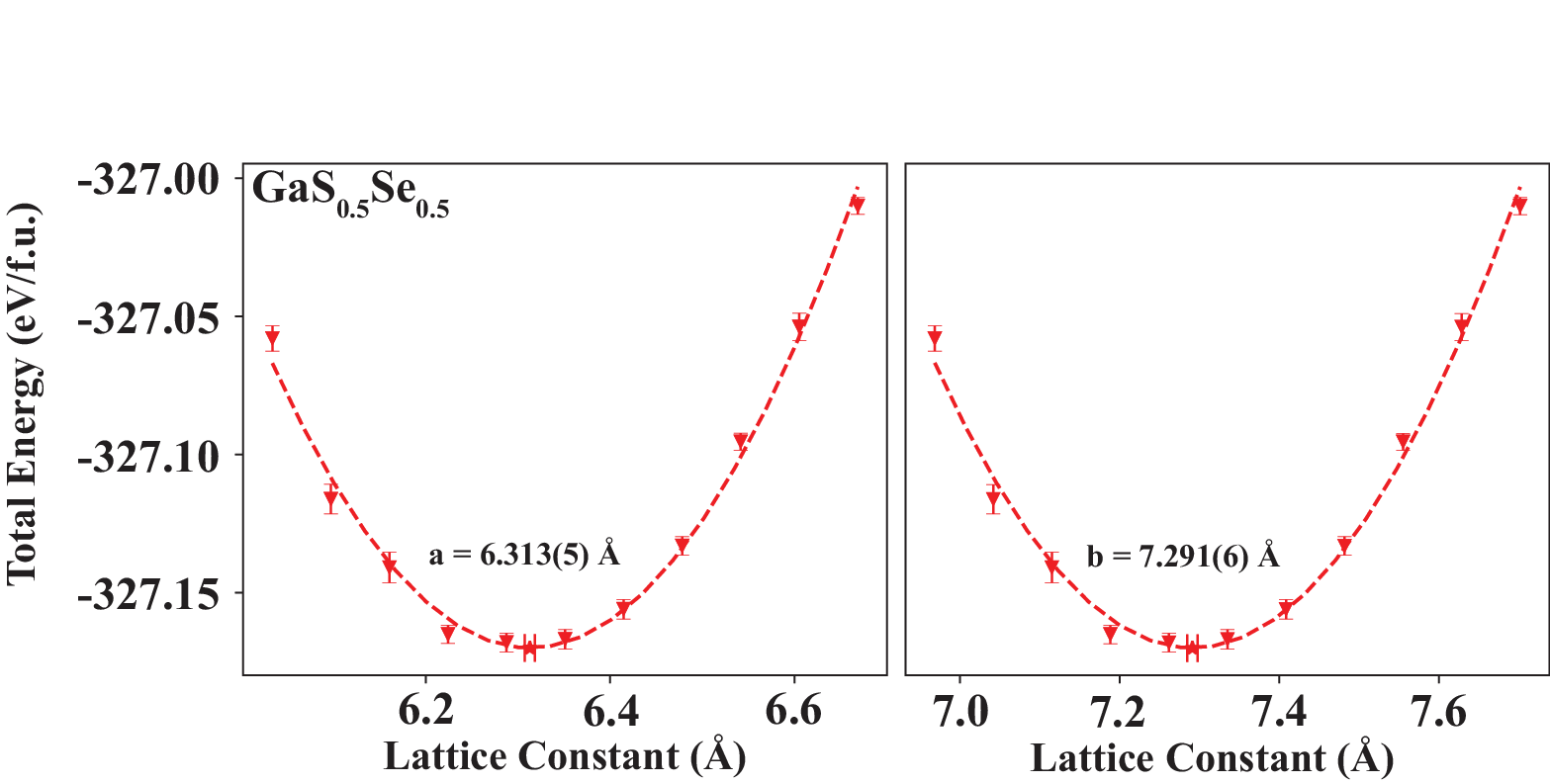}
\caption{The total energy per formula unit of 2D tetragonal GaS$_{0.5}$Se$_{0.5}$ (16 atom cell) calculated with DMC using starting wavefunctions from PBE versus the isotropically scaled lattice constants (in the $x$ direction $y$ directions respectively in each plot). The dotted lines represent the fitted curves and the minimum points and respective error bars are marked on the curves (with the value given in each of the figure insets). Geometric structure is depicted in Fig. 1 b).}
\label{gasse-0.5-geo}
\end{center}
\end{figure}

\begin{figure}
\begin{center}
\includegraphics[width=10.5cm]{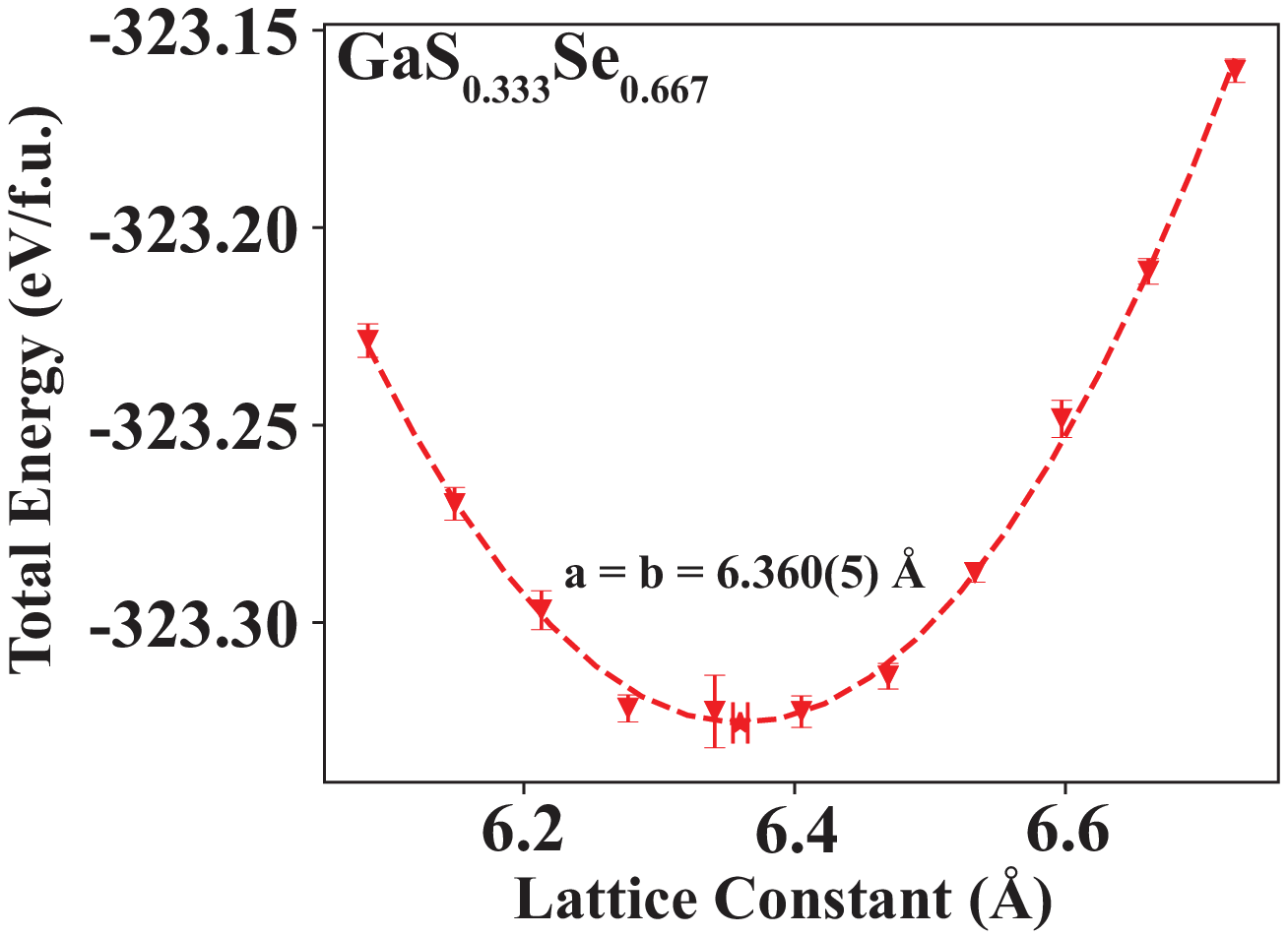}
\caption{The total energy per formula unit of 2D hexagonal GaS$_{0.333}$Se$_{0.667}$ (12 atom cell) calculated with DMC using starting wavefunctions from PBE versus the isotropically scaled lattice constant (in the $x$ and $y$ direction). The dotted line represents the fitted curve and the minimum point and respective error bar is marked on the curve (with the value given in the figure inset). Geometric structure is depicted in Fig. 1 c).}
\label{gasse-0.333-geo}
\end{center}
\end{figure}

\begin{figure}
\begin{center}
\includegraphics[width=10.5cm]{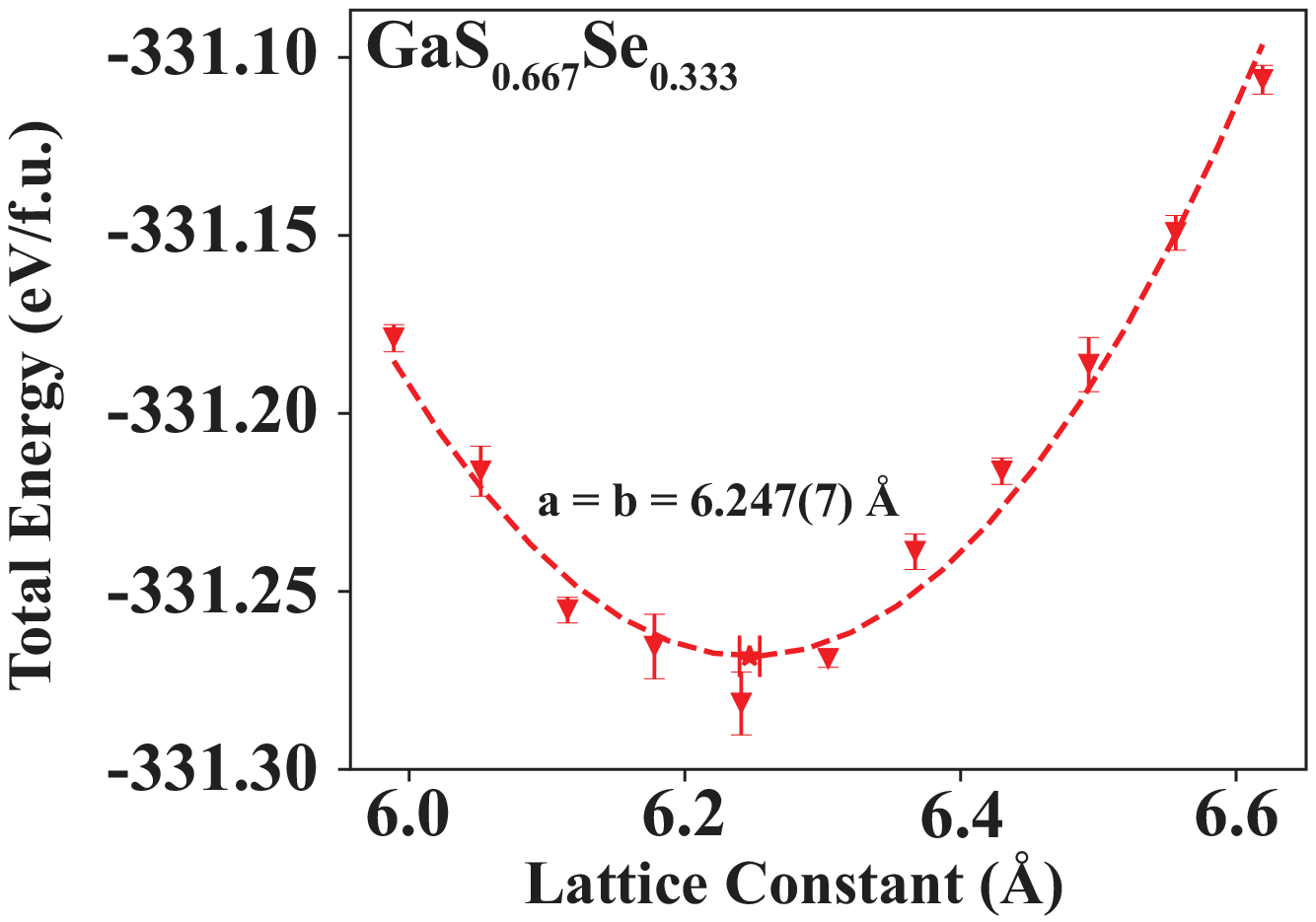}
\caption{The total energy per formula unit of 2D hexagonal GaS$_{0.667}$Se$_{0.333}$ (12 atom cell) calculated with DMC using starting wavefunctions from PBE versus the isotropically scaled lattice constant (in the $x$ and $y$ direction). The dotted line represents the fitted curve and the minimum point and respective error bar is marked on the curve (with the value given in the figure inset). Geometric structure is depicted in Fig. 1 d).}
\label{gasse-0.667-geo}
\end{center}
\end{figure}

\begin{table}[]
\begin{tabular}{c|c|c}
\hline
{ \textbf{Functional}} & { \textbf{GaSe}} & { \textbf{GaS}} \\
 & a (\AA) & a (\AA) \\
\hline
PBE                          & 3.814        & 3.637                                                        \\
\hline
PBE-D2                      & 3.745    & 3.581                                                             \\
\hline
PBE-D3                    & 3.803   & 3.625                                                              \\
\hline
SCAN                    & 3.761     & 3.599                                                      \\
\hline
SCAN+rvv10                    & 3.754   & 3.589                                                                 \\
\hline
DMC                   & 3.75(1)\cite{doi:10.1063/5.0023223}   & 3.549(3)                                                                 \\
                  & 3.74(2)\cite{doi:10.1063/5.0023223}   &                                                                  \\
\hline
Experiment                   &  3.755\cite{GaSe-optical}   &  3.587\cite{GaSe-optical}                                                                 \\
                 & 3.74\cite{GaSe-nature-synthesis}   &                                                                  \\
\hline
\hline

\end{tabular}
\caption{The calculated (using various methods) and experimental lattice constants of 2D GaS and GaSe. It is important to note that the DMC lattice constant of GaSe is 3.75(1) \AA\space when LDA and SCAN wavefunctions are used, and 3.74(2) \AA\space when PBE wavefunctions are used \cite{doi:10.1063/5.0023223}.}
\label{geo-table}
\end{table}

\begin{table}[]
\begin{tabular}{c|c|c|c|c}
\hline
{ \textbf{VMC-J3}} & { \textbf{Janus}} & { \textbf{x = 0.5}} & { \textbf{x = 0.333}} & { \textbf{x = 0.667}} \\
\hline
Janus Jastrow                          & 0                                     & 122(11)                                   & 120(13)                                    & 156(13)                                     \\
\hline
x =   0.5 Jastrow                      & 257(10)                                 & 0                                       & 33(15)                                     & 15(13)                                     \\
\hline
x =   0.333 Jastrow                    & 249(11)                                 & -10(10)                                  & 0                                         & 8(12)                                     \\
\hline
x =   0.667 Jastrow                    & 245(10)                                 & -15(11)                                  & -2(13)                                     & 0                                         \\
\hline
x =   0.875 Jastrow                    & 230(10)                                 & -47(10)                                  & -38(13)                                     & -28(12)                                    \\
\hline
{ \textbf{VMC-J2}} & {\textbf{Janus}} & { \textbf{x = 0.5}} & { \textbf{x = 0.333}} & { \textbf{x = 0.667}} \\
\hline
Janus Jastrow                          & 0                                     & 28(12)                                   & 38(13)                                     & 18(14)                                     \\
\hline
x =   0.5 Jastrow                      & 300(12)                                 & 0                                       & 27(16)                                     & -12(14)                                    \\
\hline
x =   0.333 Jastrow                    & 298(12)                                 & -10(12)                                  & 0                                         & -2(13)                                    \\
\hline
x =   0.667 Jastrow                    & 306(12)                                 & 21(12)                                   & 29(13)                                    & 0                                         \\
\hline
x =   0.875 Jastrow                    & 256(12)                                 & -33(12)                                  & -22(13)                                     & -50(13)   \\
\hline
\hline
{ \textbf{DMC-J3}} & { \textbf{Janus}} & { \textbf{x = 0.5}} & { \textbf{x = 0.333}} & { \textbf{x = 0.667}} \\
\hline
Janus Jastrow                          & 0                                     & -6(6)                                  & 2(6)                                    & 3(10)                                    \\
\hline
\hline
x =   0.5 Jastrow                      & 0(4)                                 & 0                                       & 4(7)                                     & 15(11)                                    \\
\hline
\hline
x =   0.333 Jastrow                    & -1(4)                                & -2(4)                                   & 0                                        & 17(10)                                     \\
\hline
\hline
x =   0.667 Jastrow                    & -6(5)                                & -10(6)                                  & -1(6)                                    & 0                                         \\
\hline
\hline
x =   0.875 Jastrow                    & -2(5)                                & -15(7)                                  & -2(5)                                    & 9(9)                                    \\
\hline
{ \textbf{DMC-J2}} & { \textbf{Janus}} & { \textbf{x = 0.5}} & {\textbf{x = 0.333}} & { \textbf{x = 0.667}} \\
\hline
Janus Jastrow                          & 0                                    & -3(7)                                  & -9(7)                                    & -6(7)                                    \\
\hline
x =   0.5 Jastrow                      & 3(4)                                 & 0                                       & 5(7)                                     & -9(8)                                    \\
\hline
x =   0.333 Jastrow                    & -7(5)                                & 4(9)                                   & 0                                         & 6(9)                                     \\
\hline
x =   0.667 Jastrow                    & -9(7)                                & -2(8)                                  & -2(6)                                    & 0                                         \\
\hline
x =   0.875 Jastrow                    & -8(4)                                & -5(8)                                  & -2(6)                                    & -3(7)      \\
\hline                                   
\end{tabular}
\caption{The differences (in meV/f.u.) between total energies (VMC and DMC) calculated with \textit{self-Jastrows} and \textit{shared-Jastrows} including up to two-body and up to three-body terms with the associated error bars in parenthesis. The rows represent which Jastrows are used and the columns represent each alloyed structure for which the VMC and DMC energies are calculated.}
\label{sharing}
\end{table}

The testing of our \textit{Jastrow sharing} procedure at the VMC and DMC level is tabulated in Table S3, where we show the energy difference (in meV/f.u.). At the VMC level (first and second quadrant of Table S3), we observe a higher energy difference (up to 306 meV) for the Janus structure using \textit{shared-Jastrows} with two-body and three-body interaction terms. In addition, when the Janus Jastrow parameters are used for the other subsequent structures, the energy differences are on the order of $\sim$10 meV when two-body interactions are included and on the order of $\sim$100 meV when three-body interactions are included. Due to the fact that Janus GaS$_{0.5}$Se$_{0.5}$ has an atomic environment that is slightly different than other 2D GaS$_x$Se$_{1-x}$ structures, with one face of the material being entirely S and the other being entirely Se, \textit{Jastrow sharing} at the VMC level can result in a larger energy difference between \textit{shared-Jastrows} and \textit{self-Jastrows}. A possible explanation for this is that the three-body interactions can be longer range, with electrons of Ga interacting with just S and electrons of Ga interacting with just Se on the respective surfaces (no electron mixing of S and Se).

Due to this, we reoptimized the three-body Jastrow parameters with a larger cutoff radius (3.5 \AA) and used these for DMC calculations (see Table S4). We chose 3.5 \AA \space because this is the largest value of cutoff radius that can be used for this cell size (smaller than the WS radius) and we expect to recover more three-body electron mixing interactions of S and Se. Ultimately we observe a smaller energy difference at the DMC and VMC level for the newly optimized set of three-body Jastrows with a larger cutoff radius. Although this has improved the results for the Janus structure, we cannot reuse these Jastrow parameters for the unit cells of the other alloys (x = 0.5, 0.333, 0.667), since the Jastrow cutoff for three-boy interactions is larger than the WS cell of the remaining structures. However, these newly optimized parameters could be effectively used in larger supercells.

\begin{table}[]
\begin{tabular}{c|c|c}
\hline
{ \textbf{VMC-J3}} & { \textbf{Janus}} & { \textbf{Janus}} \\
 & $r^{J3}=2$ \AA & $r^{J3}=3.5$ \AA \\
\hline
Janus Jastrow                          & -        & -                                                        \\
\hline
x =   0.5 Jastrow                      & 257(10)    & 89(11)                                                             \\
\hline
x =   0.333 Jastrow                    & 249(11)   & -98(11)                                                              \\
\hline
x =   0.667 Jastrow                    & 245(10)     & -102(10)                                                       \\
\hline
x =   0.875 Jastrow                    & 230(10)   & -116(10)                                                                 \\
\hline
{ \textbf{DMC-J3}} & {\textbf{Janus}} & {\textbf{Janus}}  \\
 & $r^{J3}=2$ \AA & $r^{J3}=3.5$ \AA \\
\hline
Janus Jastrow                          & -      & -                                                                 \\
\hline
x =   0.5 Jastrow                      & 0(4)   & 12(4)                                                             \\
\hline
x =   0.333 Jastrow                    & -1(4)   & 11(4)                                                            \\
\hline
x =   0.667 Jastrow                    & -6(5)   & 5(5)                                                                 \\
\hline
x =   0.875 Jastrow                    & -2(5)    & 9(5)                                \\
\hline
\hline
\end{tabular}
\caption{The differences (in meV/f.u.) between total energies (VMC and DMC) calculated with \textit{self-Jastrows} and \textit{shared-Jastrows} up to three-body terms with the associated error bars in parenthesis for the Janus structure (where Jastrows are optimized using two different J3 cutoff radii). The rows represent which Jastrows are used and the columns represent two different J3 cutoff radii used in the VMC and DMC calculations. }
\label{janustable}
\end{table}

For monolayer GaSe and GaS, the total energy was calculated at supercell sizes of 16, 24, 36, 48, 64 and 72 atoms. For Janus GaS$_{0.5}$Se$_{0.5}$, supercells of 16, 24, 36 and 64 atoms were used. For GaS$_{0.5}$Se$_{0.5}$ supercells of 32, 48 and 64 were used and for the GaS$_{0.333}$Se$_{0.667}$ and GaS$_{0.667}$Se$_{0.333}$ structures, supercells of 24, 36, 48 and 72 atoms were used.

\newpage 
\bibliography{si.bib}